\documentclass[journal]{IEEEtran}
\usepackage{amsmath,amsfonts}
\usepackage{algorithmic}

\usepackage{algorithm}
\usepackage{array}
\usepackage{textcomp}
\usepackage{stfloats}
\usepackage{url}
\usepackage{verbatim}
\usepackage{graphicx,cite,epsfig,amssymb,amsmath,multirow,lettrine,flushend,extarrows,makecell}
\usepackage{textcomp}
\usepackage{lipsum}
\usepackage{amsthm}
\usepackage{siunitx}
\usepackage{amsmath}
\usepackage{multicol}
\usepackage{multirow}
\usepackage{mathtools}
\usepackage{color}
\usepackage{bm}
\usepackage{indentfirst}
\usepackage{nomencl}
\usepackage{booktabs}
\usepackage{cite}
\usepackage{mathtools}
\usepackage{accents}
\usepackage{tikz}
\usepackage[numbers,sort&compress]{natbib}
\usepackage{subcaption}
\usepackage{amssymb}

\usepackage{booktabs}
\usepackage{array}
\usepackage{makecell}

\newtheorem{theorem}{Theorem}

\newtheorem{remark}{Remark}

\newcommand{\A}{\bm{A}}
\newcommand{\B}{\bm{B}}

\newcommand{\T}{\bm{T}}
\newcommand{\R}{\bm{R}}
\newcommand{\aaa}{\bm{a}}
\newcommand{\q}{\bm{q}}
\newcommand{\dd}{\bm{d}}
\newcommand{\m}{\bm{m}}
\newcommand{\w}{\bm{w}}
\newcommand{\soz}{\bm{{\rm soz}}}
\newcommand{\soc}{{\rm soc}}

\usepackage{booktabs} 

\begin{document}

\title{An Equivalent and Unified Virtual Battery Modeling Framework for  Flexibility Characterization of Building HVAC Systems}

\author{ 
		Qi~Zhu, 
		Yu~Yang,~\IEEEmembership{Member,~IEEE,}
		Liang~Yu,~\IEEEmembership{Senior Member,~IEEE,}
		Qing-Shan Jia, ~\IEEEmembership{Senior Member,~IEEE,}
        Costas~J.~Spanos,~\IEEEmembership{Fellow,~IEEE,}
		Xiaohong~Guan,~\IEEEmembership{Life Fellow,~IEEE}
		\thanks{This  work  was  supported by the National Natural Science Foundation of China  (62403373, 62192752).}
		\thanks{Q. Zhu and Y. Yang are with School of Automation Science and Engineering, Xi’an Jiaotong University, Shaanxi, Xi’an 710049, China  (e-mail:  zhuqi@stu.xjtu.edu.cn, yangyu21@xjtu.edu.cn).
			\textit{Y. Yang is the corresponding author.}
		}
	\thanks{L. Yu is with the College of Automation, Nanjing University of Posts
	and Telecommunications, Nanjing 210023, China (e-mail: liang.yu@njupt.edu.cn).}
\thanks{Q. -S. Jia is with the Center for Intelligent and Networked Systems,
	Department of Automation, BNRist, Tsinghua University, Beĳing 100084,
	China (e-mail: jiaqs@tsinghua.edu.cn).}
		\thanks{X. Guan is with School of Automation Science and Engineering, Xi’an Jiaotong University, Shaanxi, Xi’an 710049, China, and also 	
			with the Center for Intelligent and Networked Systems,
			Department of Automation, Tsinghua University, Beijing 100084, China (e-mail: xhguan@tsinghua.edu.cn).}
		\thanks{C. J. Spanos is with the Department of Electrical Engineering and 	Computer Sciences, University of California, Berkeley, CA, 94720 USA  (email: spanos@berkeley.edu).} 
	}

\markboth{IEEE Transactions on Smart Grid,~Vol.~XX, No.~XX, Month~2025}%
{An Unified Virtual Battery Modeling Framework for  Flexibility Characterization of Building HVAC Systems}


\maketitle
\begin{abstract}
The  heating, ventilation and air-conditioning (HVAC) system dominates building's energy consumption and meanwhile exhibits substantial operational flexibility that can be exploited for providing grid services.
However, the goal is largely hindered by the difficulty to characterize the system's operating flexibility  due to the complex building thermal dynamics, system operating limits and human  comfort constraints.
To address this challenge, this paper develops an unified virtual battery (VB) modeling framework for characterizing the operating flexibility of both single-zone and multi-zone building HVAC systems, enabling flexible buildings to function like virtual batteries. 
Specifically,  a physically meaningful  representation state is first identified to represent  building thermal conditions under thermal comfort constraints and a VB model is then established for characterizing the operating flexibility of single-zone HVAC systems. We subsequently extend the VB modeling framework to multi-zone HVAC systems and establish a set of zone-level VB models to characterize the building's zonal operating flexibility.  We further develop a systematic method to aggregate the VB models into a low-order and low-complexity aggregated VB model, significantly reducing model and computational complexity. 
We demonstrate the   VB model through  demand response (DR) applications and conclude that the VB model can well capture the operating flexibility of building HVAC systems and enable effective DR participation. The DR strategies obtained from the VB model can be efficiently decomposed to  zone-level control inputs for maintaining human thermal comfort while achieving near-optimal operation cost.  
\end{abstract}


\section{Introduction}

\IEEEPARstart{B}{uildings} account for over 35\% of global energy consumption, while the heating, ventilation, and air conditioning (HVAC) systems for space conditioning contribute about 40\%–50\% of buildings' total energy use  \cite{IEA_Energy_Efficiency_2025}. Particularly, this proportion is continuing increasing and projected to grow  by more than 50\% by 2050 \cite{wijesuriya2025enhancing}. As the main energy sector, buildings' energy consumption shows substantial operating flexibility due to the buildings' thermal inertia and wide comfortable temperature ranges of occupants. Specifically, 
the thermal energy generated by HVAC units  can be temporally and  bidirectionally modulated in response to power grid operating requirements. 
Building  HVAC systems  have been recognized as one of the most important flexible resources for alleviating the flexibility shortage of modern power grids  \cite{GEB_WholeBuilding_2019} and providing a wide range of grid services, including demand response \cite{kou2021model}, peak shaving \cite{anuntasethakul2021design}, frequency regulation \cite{qureshi2018hierarchical}, etc.

The exploitation of operating flexibility of building HVAC systems rely on tractable models for flexibility characterization. Considerable efforts have been made for developing models for capturing building thermal dynamics under HVAC control inputs. The existing models can be broadly  classified into white-box, black-box and gray-box models (see \cite{drgovna2020all} for a comprehensive review).  White-box models  depend on professional simulators, such as EnergyPlus \cite{energyplus} and TRNSYS \cite{trnsys}, to capture the detailed physical principles and  emulate real operating scenarios. 
Back-box models generally employ common expressive functional modules, such as neural networks \cite{smarra2018data}, linear regression models \cite{lu2021data},  to capture the complex mappings  between HVAC control inputs, external thermal disturbances and building thermal dynamics. The models often involve a large number of parameters and require high-quality operating data for model parameter identification. In contrast, gray-box models generally rely on simplified physical principles, such as energy and mass conservation, heat transfer and convection principles, for mathematical models establishment, enjoying the benefit of  limited model parameters and high model generality.  Among  the existing models, the resistor–capacitor (RC) models that employ  thermal resistance and capacitance to characterize heat transfer efficiency and the thermal inertia of building components \cite{deng2010building, maasoumy2011model},  
  have been most widely used for control-oriented applications (see \cite{yang2021distributed,yang2020hvac} for examples) due to their high interpretability.

However, most of the existing models for thermal dynamics modeling for building HVAC system  are not suitable for characterizing the system's  operating flexibility. 
They are often subject to high-dimensional nonlinear thermal dynamics, operating limits and  human comfort constraints.   When considering the unlock of building operating flexibility for mangy high-level energy management or scheduling tasks,  a low-order and low-complexity model that provides the  energy scheduling boundaries  would be preferred rather than a detailed zone-level control models that are computationally intensive. 
 
 \begin{figure*}[t]
 	\centering
 	\includegraphics[width=0.93\textwidth]{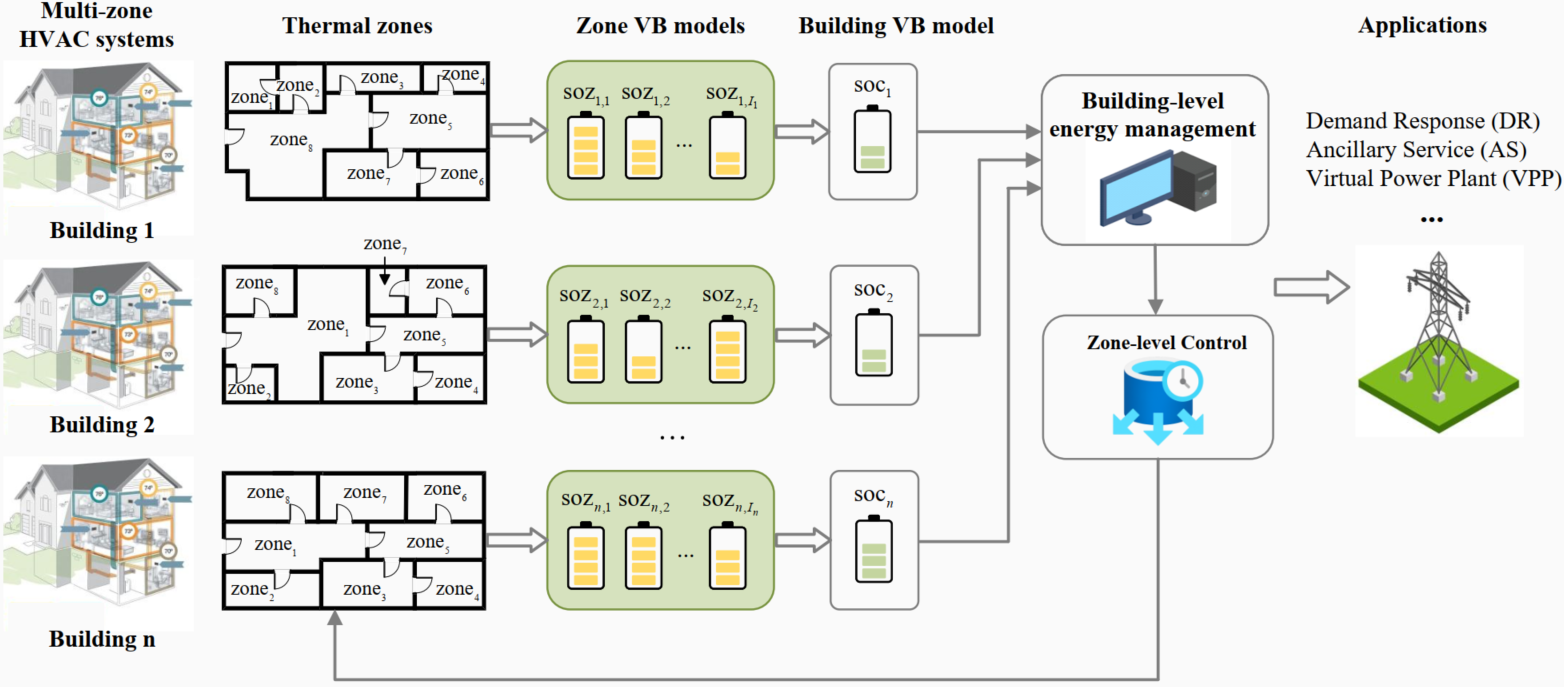}
 	\caption{VB model for characterizing operating flexibility of building HVAC systems}
 	\label{fig:main_figure}
 \end{figure*} 

To this end, many researches has focused on developing  low-order and low-complexity models for characterizing the operating flexibility of building  HVAC systems. For single-zone  HVAC systems (often used in residential buildings), numerous studies have  developed virtual battery (VB) models for characterizing   their operating flexibility that show high interpretability and are easy to be coordinated.  Whereas the problem is much more complicated and challenging  with multi-zone HVAC systems (often used in commercial buildings) that  show quite different operating behaviors and admit high-dimensional nonlinear state-space formulations. 
 To address this challenge, several studies have  relied on  data-driven  approaches to capture the aggregated operating flexibility of 
multi-zone HVAC systems \cite{cui2024data, cui2025dimension}. Whereas the model constructions are very sophisticated and lack of theoretical guarantee. A number of works have developed optimization-based methods to approximate the system's operating flexibility by virtual battery representation, but depend on solving a sequence of comprehensive optimization problems for model parameter identification \cite{hao2017optimal}. 
From the literature, we find that characterizing the operating flexibility for building HVAC systems   is a nontrivial problem and  a systematic approach is still lacking. 

To fill the gap, this paper  develops a virtual battery framework to characterize the operational flexibility for  both  single-zone and multi-zone building HVAC systems. The VB models are established from the well-known RC formulations via equivalent mathematical transformations, thus with readily available model parameters and  theoretic guarantee.  
Particularly, the  VB models  provide high interpretability and enable flexible buildings to function like virtual batteries when interact with the power grid operators. A schematic illustration of our idea  is presented in Fig. \ref{fig:main_figure}. The VB models are applicable to a wide range of building-level and cluster-level energy management and scheduling tasks while leveraging  buildings' operating flexibility. \textbf{The main contributions of this paper are summarized as follows}:
\begin{itemize}
	\item We propose a compact and physically meaningful \emph{characterization state} to  represent building thermal conditions under  thermal comfort constraints, 
	and further establish a VB model for characterizing the operational flexibility of single-zone building HVAC systems.
	\item We extend the VB modeling framework to multi-zone building HVAC systems and establish a set of zone-level VB models to capture  the heterogeneous zonal thermal dynamics and operating flexibility.
	\item We further develop a systematic  method for aggregating the zone-level VB models into a low-order and low-complexity aggregated VB model, significantly reducing model and computational complexity.
\end{itemize} 

We demonstrate the effectiveness of VB models  through demand response (DR) applications. The results show that the low-order and low-complexity  VB models can well capture  the thermal dynamics of  buildings under varying HVAC control and can enable effective DR participation.  
The rest of this paper is summarized: In Section II, we review the related works on 
the operating flexibility characterization and modeling  of building HVAC systems; In section III and IV, we introduce the VB models for  single-zone and multi-zone  HVAC systems. In Section V, the VB models are validated and tested for DR applications. In Section VI, we conclude this paper and discuss future works.

\renewcommand{\arraystretch}{1.3} 
\begin{table*}[htbp] 
	\centering
	\caption{Existing works of flexibility characterization for building HVAC systems}
	\label{tab:table_literature}
	\begin{tabular}{clllllll} 	
		\toprule
		Systems                
		& Methods                  
		& Models                          
		& \makecell[c]{Method\\complexity} 
		& \makecell[c]{Theoretic \\ Guarantee}      
		& Year 
		& Paper  \\
		\hline       
		\makecell[c]{ Multi-zone HVAC \\ Single-zone HVAC}  & Optimization-based     & Generalized Battery Model & High     & No         & 2017         & \cite{hao2017optimal} \\
		\hline
		Single-zone HVAC       & Optimization-based     & Polytope                                 & High     & No        & 2017, 2025   & \cite{zhao2017geometric, mukhi2025aggregate} \\
		Single-zone HVAC         & Data-driven        & Linear regression                      & Low     & No         & 2021, 2024     & \cite{lu2021data, hou2024privacy}  \\[4pt]
		Single-zone HVAC         & Model-based        & Virtual Battery Model                      & Low     & Yes         &2017, 2019     & \cite{song2017thermal, song2019hierarchical}  \\[4pt]
		Single-zone HVAC         & Model-based        & Virtual Battery Model                       & Low     & Yes         & 2025     & \cite{zhu2025multi}  \\[4pt]	
		\hline
		Multi-zone HVAC     & Data-driven            & Virtual Storage Model                  & High     & No         & 2024     & \cite{cui2024data}  \\[4pt]
		Multi-zone HVAC     & Data-driven           & Neural Network                      & High     & No         & 2024     & \cite{cui2025dimension}  \\[4pt]
		Multi-zone HVAC     &  Model-based           & Virtual Battery Model                   & Low      & No         & 2022         & \cite{abbas2022evaluation} \\
		Multi-zone HVAC     & Model-based           & 2R2C                       & Low     &  No         & 2019,2021     & \cite{guo2019identification, guo2021aggregation}  \\[4pt]
		\hline 
		\noalign{\vskip 2pt}
		\makecell[c]{ \textbf{Multi-zone HVAC} \\ \textbf{Single-zone HVAC}} & \makecell[l]{\textbf{Model-based}} & \textbf{Virtual Battery  Model}   &  \textbf{Low} & \textbf{Yes}  &  \textbf{2025} & \textbf{This paper}\\
		\bottomrule
	\end{tabular}
\end{table*}

\section{Literature}

A substantial body of researches has focused on the operating flexibility characterization of building HVAC systems.  Existing works as well as this paper are contrasted in Table \ref{tab:table_literature},  including the types of systems considered (\texttt{Systems}), the  approaches used to build the models (\texttt{Methods}), the resulted model formulations (\texttt{Models}), the computational  complexity of the model parameters identification (\texttt{Model Complexity}),
 and theoretical guarantee of model accuracy  (\texttt{Theoretic Guarantee}).

The existing works mainly focused on  single-zone  and multi-zone HVAC, respectively. This is mainly because the two types of systems  operate quite differently and show different operating behaviors. The former typically operates with constant ventilation rates and admits linear state-space models, while the later generally operates with variable ventilation rates and depend on high-dimensional and nonlinear state-space model for capturing spatially-temporally coupled thermal dynamics.  The approaches can be classified into  optimization-based, data-driven and model-based.  
Optimization-based methods are generally about  identifying a group of parameters for  user-defined characterization models, including virtual battery models \cite{hao2017optimal} or  polytope models \cite{zhao2017geometric, mukhi2025aggregate},  to approximate the operating flexibility of building HVAC systems.    These approaches are usually computationally intensive  as  a sequence of  comprehensive optimization problems are required to be solved for model parameters identification. 
Data-driven approaches generally share the idea of using  functional  modules to approximate  the buildings'  thermal dynamics, including piece-wise linear  (PWL) functions \cite{cui2024data}, linear regression models  \cite{lu2021data, hou2024privacy} and neural networks \cite{cui2025dimension}, etc.  Data-driven approaches are advantageous in handling the complex system dynamics but the resulted models often lack generality and theoretic guarantee. 
In contrast, model-based methods generally proposed to establish  characterization  models from  existing building thermal dynamic models, typically the RC formulations,  via analytic or mathematical  transformations (see \cite{song2017thermal, song2019hierarchical} for examples). They often provide interpretability and theoretic soundness.

This above methods have  led to a  set of characterization models, including  virtual energy storage (VES) or virtual battery (VB) models,  polytopes, linear regression models and neural networks, among others.  
The VES and VB models have been widely employed mainly due to their natural capability for flexibility characterization and  easy to be coordinated with other types of energy storage systems.  For single-zone HVAC, a number of  studies has developed  VB models for characterizing  the system's  operating  flexibility  \cite{song2017thermal, song2019hierarchical, zhu2025multi}.  Whereas the issue with multi-zone HVAC systems has been largely unresolved. This is mainly due to the systems' complicated and quite different operating behaviors, leading to  high-dimensional and  nonlinear state-space models. This  makes the development of low-order models for capturing and aggregating the zonal flexibility mathematically complicated. 
Notably, \cite{hao2017optimal, abbas2022evaluation} proposed to characterize the flexibility of multi-zone HVAC by VES models  using optimization-based techniques to identify  model parameters.   \cite{cui2024data} proposed to train  PWL  functions to approximate the non-linear and complicated  thermal dynamics for VES modeling.  Slightly different, \cite{cui2025dimension}  proposed to learn a low-order model in the latent space using machine learning techniques to overcome the model complexity. \cite{guo2019identification, guo2021aggregation} have derived an aggregated RC model  from the original RC formulations for  multi-zone HVAC systems,  but face the difficulty to handle model parameters as they are closely coupled with zone-level state and decision variables. 

From the literature, we find that many efforts have been made for flexibility characterization of building  HVAC systems, driven by  the growing demand to unlock the enormous  flexibility  for supporting modern power operation.  However, the problem is still largely unresolved and there still lacks a systematic framework account  for  both single-zone and multi-zone HVAC systems. Moreover, existing  works  mostly relied on optimization-based and data-driven approaches  for model constructions, suffering from high computational complexity in model parameters identification,  limited generality and theoretic guarantee.

\section{Virtual Battery Model for Single-zone HVAC}
For single-zone HVAC systems that are  typically  used in residential buildings, the widely used  resistor-capacitor (RC) models take the formulation of \cite{hao2017optimal, qiu2023federated}.
\begin{subequations} 
	\begin{align}
		&C^{\rm th}_n \left( T_{n}(k+1) - T_{n}(k) \right) = \frac{T^{\rm out}(k) \!- \!T_{n}(k)}{R^{\rm th}_{n, o}} \Delta k  \notag  \\
		& \quad \quad \quad \quad \quad \quad  - \eta_{n}^{\rm HVAC} q_{n}^{\rm HVAC}(k) \Delta k + Q_{n}^{\rm dist}(k)\Delta k,  \label{eq:AC_dynamics}\\
		& q_{n}^{\rm HVAC, \min} \leq q_{n}^{\rm HVAC}(k) \leq q_{n}^{\rm HVAC, \max}, \label{eq:operating_limits} \\
		& T_n^{\rm set} - \delta_n \leq T_{n}(k) \leq T_n^{\rm set} +  \delta_n, \quad \forall k \in \mathcal{K}. \label{eq:thermal_comfort} 
	\end{align} 
\end{subequations}
where $n$ and $k$ are building and time index. Equations \eqref{eq:AC_dynamics} represent zone thermal dynamics.  
 $C^{\rm th}_{n}$ is zone  thermal capacity, $T_{n}(k)$ is zone  air temperature, $T^{\rm out}(k)$ is outdoor temperature, $T^{\rm sup}_n$ denotes the  setpoint temperature of supplied airflow,  $R^{\rm th}_{n,o}$ represents the thermal resistance
	between the zone and the outside,  $Q^{\rm dist}_{n,i}(k)$ denotes the thermal disturbances caused by occupants, electrical devices and solar radiations, etc.
$q_{n}^{\rm HVAC}(k)$ denotes the controllable  power consumption of the HVAC unit, with $\eta_{n}^{\rm HVAC}$ denoting the chiller's  coefficient of performance (COP).   Constraints \eqref{eq:operating_limits} specify  the feasible range of HVAC power consumption, with $q_{n}^{\rm HVAC, \min}$ and $q_{n}^{\rm HVAC, \max}$ denoting the lower and upper limits.  
   Constraints \eqref{eq:thermal_comfort} model human thermal comfort, with  $T_{n}^{\mathrm{set}}$ denoting the ideal zone  temperature  setpoint and $\delta_{n}$ indicating the allowable upward and downward deviation.  Note that the single-zone HVAC systems that typically operate in constant ventilation rates,  admit  linear state-space models. 
The zone thermal dynamics \eqref{eq:AC_dynamics} can be stated in a compact format:
\begin{equation} 
	T_{n}(k+1) = a_n T_{n}(k) - b_n q_{n}^{\rm HVAC}(k)  + d_{n}(k), \forall k \in \mathcal{K}. \\
\end{equation}
where we have $a_n = 1 - \Delta k/(C_n^{\rm th} R_{n, o}^{\rm th})$, $b_n = \eta^{\rm HVAC}  / C_n^{\rm th} \Delta k$ and $d_{n}(k) = T^{\rm out}(k)\Delta k /(C_n^{\rm th} R_{n,o}^{\rm th}) + Q_{n}^{\rm dist}(k) \Delta k/C_n^{\rm th}$.

\subsection{Characterization State} Before establishing the VB model, we first define the following state-of-zone (soz)  as the \emph{characterization state} of   zone thermal conditions under thermal comfort constraints:  
\begin{equation} \label{eq:soz}
	{\rm soz}_{n}(k) = \frac{T_{n}^{\rm set} - T_{n}(k)}{\delta_{n}}, \quad \forall k \in \mathcal{K}.
\end{equation}
Particularly, the soz can be viewed as ``State-of-Comfort” of the zone.  
When the zone  is controlled  within the comfortable temperature range, it follows that ${\rm soz}_{n}(k) \in [-1, 1],  k \in \mathcal{K}$.
Meanwhile, the soz can be viewed as ``State-of-Charge” of the zone,  with the upper and lower limits $1$ and $-1$ denoting a full and an empty zone state. 
Particularly, when the zone reaches  lower temperature limit, it can not consume any energy further (i.e., charging) and we have  a full zone indicated by  ${\rm soz } = 1$. 
Oppositely, when the zone reaches upper temperature limit,  the zone can not reduce any energy consumption (i.e., discharging) and we have an empty  zone indicated by ${ \rm soz } = -1$. The soz can be interpreted as the zone thermal conditions and is also the basis to establish a VB model from the RC formulation.

\subsection{VB Model for Single-zone HVAC}
For the operation of  HVAC system, an ideal setting is to maintain zone temperature  at the setpoints. We refer to this control policy as \texttt{Baseline control}, corresponding to the following steady state equations by  \eqref{eq:AC_dynamics}:
\begin{equation} \label{eq:AC_Baseline}
	T_{n}^{\rm base}(k\!+\!1) \!\!=\!\! a_n T_{n}^{\rm base}(k) \!\!-\!\! b_n q_{n}^{\rm HVAC, base}(k) \!\!+\!\! d_{n}(k), \forall k \in \mathcal{K}. \\
\end{equation}

By subtracting \eqref{eq:AC_dynamics} from \eqref{eq:AC_Baseline} and dividing $\delta_{n}$, we have
\begin{equation}
	\begin{split}
	&\frac{T_{n}^{\rm base}(k\!+\!1) \!\! - T_{n}(k\!+\!1)}{\delta_n}=\! a_n \frac{ T_{n}^{\rm base}(k) -T_{n}(k) }{\delta_n} \!\!\\
	&  \quad \quad + b_n ( q_{n}^{\rm HVAC}(k) - q_{n}^{\rm HVAC, base}(k) ), \quad \forall k \in \mathcal{K}. 
	\end{split}
\end{equation}

By introducing  the characterization state \eqref{eq:soz}, we  have 
\begin{equation} \label{eq:AC_soz_dynamics}
	\begin{split}
		{\rm soz}_{n}(k+1) = a_n {\rm soz}_{n}(k) +  P^{\rm ch/dis}_{n}(k)
	\end{split}
\end{equation}
where $P^{\rm ch/dis}_{n}(k) \triangleq {b_n}/{\delta_n}\left(q_{n}^{\rm HVAC}(k) - q_{n}^{\rm HVAC, base}(k) \right)$ can be viewed as  the net charging power of a VB model.

By further encoding the operating and thermal comfort constraints \eqref{eq:operating_limits}-\eqref{eq:thermal_comfort}  with $P^{\rm ch/dis}_{n}(k)$ and ${\rm soz}_{n}(k)$,  we can obtain the following  VB models for single-zone building HVAC systems:
\begin{equation} \label{eq:AC_VB}
	\begin{split}
		& {\rm \textbf{VB model for single-zone HVAC:}} \\
		& \begin{cases}
			{\rm soz}_{n}(k+1) = a_n {\rm soz}_{n}(k) +  P^{\rm ch/dis}_{n}(k), \\
			-1 \leq {\rm soz}_n(k) \leq 1, \\
			P^{\rm ch/dis}_{n}(k)  \!\geq\!q_{n}^{\rm HVAC, \min} - q_{n}^{\rm HVAC, base}(k), \\ 
			P^{\rm ch/dis}_{n}(k) \!\leq\! q_{n}^{\rm HVAC, \max} - q_{n}^{\rm HVAC, base}(k), \forall k \in \mathcal{K}.
		\end{cases}
	\end{split}
\end{equation}

 For a single-zone HVAC system, the related energy consumption at time step $k$ is $q_{n}^{\rm HVAC}(k) \Delta k$, which can be   equivalently expressed by 
\begin{equation}
 Q^{\rm tol}_n(k) \!\!=\!\! (P^{\rm ch/dis}_{n}(k)\delta_{n}/b_n \!+\! q_{n}^{\rm HVAC, base}(k) )\Delta k, \forall k \in \mathcal{K}. 
\end{equation}
with  VB model \eqref{eq:AC_VB}.

\section{Virtual Battery Models for Multi-zone HVAC}

This section discusses  extension of the VB modeling framework to multi-zone building HVAC systems. This problem is non-trivial as the systems show much more complicated operating behaviors than single-zone cases. We refer the readers to \cite{yang2020hvac, lin2015experimental} for more details.

\subsection{RC models for multi-zone HVAC systems}
For multi-zone HVAC systems typically deployed in commercial buildings, the general RC  formulations are \cite{yang2020hvac, wang2017distributed}:  	
\begin{subequations} 
	\begin{align}
& C^{\rm th}_{n,i}\big(T_{n,i}(k\!+\!1) \!-\!  T_{n,i}(k)\big) \!\!=\!\!\!\!\sum_{j \in \mathcal{I}_n(i)}\!\!\! \frac{T_{n,j}(k) \!-\! T_{n,i}(k)}{R^{\rm th}_{n,ij}} \Delta k \notag \\
&\! + \!\frac{T^{\rm out}(k) \!\!-\!\! T_{n,i}(k)}{R^{\rm th}_{n,oi}} \Delta k \!+\! c_p m_{n,i}(k)\left(T^{\rm sup}_n \!\!-\!T_{n,i}(k)\right) \Delta k \notag \\
&  \quad+ Q^{\rm dist}_{n,i}(k) \Delta k, \quad \quad \quad \quad \quad \quad \quad \quad  \forall i \in \mathcal{I}_n, k \in \mathcal{K}.  \label{eq:zone_dynamics}\\
& m_{n,i}^{\min} \leq m_{n,i}(k) \leq m_{n,i}^{\max}, \quad \quad \quad \quad \quad \forall i \in \mathcal{I}_n, k \in \mathcal{K}.   \label{con:zone_air_flow}\\
& T_{n,i}^{\rm set} -\delta_{n,i}  \leq T_{n,i}(k) \leq T_{n,i}^{\rm set} + \delta_{n,i}, \quad \forall i \in \mathcal{I}_n, k \in \mathcal{K}. \label{con:zone_comfort}
	\end{align}
\end{subequations}
where $n, i, k$  are building, zone and time index. We use $\mathcal{I}_n$ and $I_n$ to  denote the set and number of zones involved. Equations \eqref{eq:zone_dynamics} represent the spatially-temporally coupled zone thermal dynamics.  $R_{n,ij}$ denotes the thermal resistance between the adjacent zone $i$ and zone $j$.   $\mathcal{I}_n(i)$ denotes the set of  zones adjacent to zone $i$.  $c_p$ is the specific heat of air. $m_{n,i}(k)$  denotes the zone airflow rates to be controlled. The other notations follow the single-zone case. Constraints~\eqref{con:zone_air_flow} specify the feasible ranges of zone airflow rates, with $m_{n,i}^{\min}$ and $m_{n,i}^{\max}$ denoting the minimum and maximum values determined by the VAV boxes operating limits. Constraints~\eqref{con:zone_comfort} model zone thermal comfort constraints. Similarly, the zone thermal dynamics \eqref{eq:zone_dynamics_compact} can be presented  in a compact format:
\begin{equation} \label{eq:zone_dynamics_compact}
	\begin{split} 
		&T_{n,i}(k \!+\! 1) \! \!=\!\! a_{n,ii} T_{n,i}(k) \!\!+\!\!\!\!\!\!\!\sum_{j \in \mathcal{I}_n(i)}\!\!\!\!\! a_{n,ij}T_{n,j}(k) \!\!+\!\! a^{\rm out}_{n,i}T^{\rm out}(k)\\
		& \quad \quad \quad - b_{n,i} q_{n,i}(k) + d_{n,i}(k), \\
		& q_{n,i}(k) = c_p m_{n,i}(k)(T_{n,i}(k) \!\!-\!\! T^{\rm sup}_n), \forall i \in \mathcal{I}_n, \forall k \in \mathcal{K}.
	\end{split}
\end{equation}
where the parameters $a_{n,ii}, a_{n,ij}, a^{\rm out}_{n,i}, b_{n,i}, d_{n,i}(k)$  are 
\begin{equation*}
	\left\{
	\begin{aligned}
		&a_{n,ii}\!=\!1 \!-\! {\Delta k}/{(R^{\rm th}_{n,io} C_{n,i}^{\rm th})} \!\!-\!\!\!\!\!\!\!\sum_{j \in \mathcal{I}_n(i)}\!\!\!\!{\Delta k}/{(R^{\rm th}_{n,ij} C_{n,i}^{\rm th})},\!\!\!\!\!&& \forall i \in \mathcal{I}_n. \\
		&a_{n,ij} = {\Delta k}/{(C_{n,i}^{\rm th} R^{\rm th}_{n,ij})}, && \forall i, j \in \mathcal{I}_n. \\
		&a^{\mathrm{out}}_{n,i} ={\Delta k}/{(R^{\rm th}_{n,io} C_{n,i}^{\rm th})}, && \forall i \in \mathcal{I}_n. \\
		&b_{n,i} ={\Delta k}/{C_{n,i}^{\rm th}}, && \forall i \in \mathcal{I}_n. \\
		&d_{n,i}(k) = {Q_{n,i}^{\rm dist}(k) \Delta k}/{C_{n,i}^{\rm th}}, \quad \quad \quad \quad ~ \forall i \in \mathcal{I}_n,&&k \in \mathcal{K}.
	\end{aligned}
	\right.
\end{equation*}

By stacking \eqref{eq:zone_dynamics_compact} across the thermal zones, we have  the  zone thermal dynamics of  matrix format:
\begin{equation} \label{eq:optimize}
	\begin{split}
		& \T_n(k\!+\!1) \!\!=\!\! \A_n \T_n(k)\! \!-\!\!\B_n \q_n(k) \!+\! \aaa_n^{\rm out} \T^{\rm out}(k) \!\!+\!\! \dd_n(k), \\
		& \q_n(k) = c_p \m_n(k)\left(\T_n(k) - \mathbf{1}_n T^{\rm sup}_n \right), \forall k \in \mathcal{K}.
	\end{split}
\end{equation}
where we have 
\begin{equation*}
	\begin{split}
		 \A_n  & \!=\! \big[a_{n, ij}\big]_{i,j \in \mathcal{I}_n} \in R^{I_n \times I_n},  \\
		 \B_n   & \!=\! \mathrm{diag}(b_{n,1}, b_{n,2}, \dots, b_{n,I_n}) \in R^{I_n \times I_n},\\
		 \mathbf{1}_n & \!=\! [1, 1, \cdots, 1]^{\mathsf T} \in R^{I_n}. 
	\end{split}
\end{equation*}

The energy consumption of multi-zone HVAC systems consists of cooling power and  fan power, which are  \cite{yang2020hvac,  wang2017distributed} 
\begin{equation} \label{eq:cooling_power_fan_power}
\begin{split}
P_{\rm cooling}(k)  = & c_p (1 - d_r) \sum_{i \in \mathcal{I}_n} m_{n,i}(k)(T^{\rm out}(k) - T^{\rm sup}_n)  \\
& + c_p d_r \sum_{i \in \mathcal{I}_n} m_{n,i}(k) (T_{n,i}(k) - T^{\rm sup}_n),   \\ 
P_{\rm fan}(k) = &  \kappa_f \left( \sum\nolimits_{i \in \mathcal{I}_n }m_{n,i}(k) \right)^2, \forall k \in \mathcal{K}. 
\end{split}
\end{equation}
where $d_r \in [0, 1]$ denotes the fraction of return air  to AHU. The cooling power is divided into two parts: cooling the outdoor fresh air and  the return air entering the AHU to the setpoint. The fan power  is calculated  by total zone airflow rates. The cooling power is generated by the chilled water pumped from the chiller and thus its  electrical consumption is determined by affected the chiller’s COP. Collectively, the system's total energy consumption can be expressed as
\begin{equation} \label{eq:AHU_power}
\begin{split}
	Q_n^{\rm tol}(k)\! =\! P_{\rm cooling}(k)/{\rm COP} \Delta k  + P_{\rm fan}(k) \Delta k, \forall k \in \mathcal{K}. 
\end{split} 
\end{equation}

\subsection{VB Models for Multi-zone HVAC}
It is clear from \eqref{eq:zone_dynamics}-\eqref{con:zone_comfort} that the operation of multi-zone HVAC systems are governed by   high-dimensional  nonlinear state-space models, which are in contrast to the linear state-space models with single-zone cases. Particularly, the numbers of decision variables, operating constraints and the non-linear thermal dynamic equations  grow  with the number of thermal zones. 
As a result,  the energy management tasks of   multi-zone HVAC systems relying on the  RC formulations often face substantial computational complexity. 
To this end, this section extends the VB modeling framework to the multi-zone cases. We adopt the same \emph{characterization state}  to represent zone thermal conditions under thermal comfort constraints, i.e., ${\rm soz}_{n, i}(k) =(T_{n, i}^{\rm set} - T_{n, i}(k))/\delta_{n, i}, \forall i \in \mathcal{I}_n,  k \in \mathcal{K}$.  Similarly, we refer to maintaining  zone temperature  at their setpoints as \texttt{Baseline control}, leading to the following steady state equations by \eqref{eq:optimize}. 
\begin{equation} \label{eq:baseline}
	\begin{split}
		& \T^{\rm set}_n(k\!+\!1) \!\!=\!\! \A_n \T^{\rm set}_n(k) \!-\!\B_n \q^{\rm base}_n(k) \!+\! \aaa^{\rm out}_n \T^{\rm out}(k) \!\!+\!\! \dd_n(k), \\
		& \q^{\rm base}_n(k) \!=\! c_p \m_n(k)\left(\T^{\rm set}_n(k) - \mathbf{1}_{n} T^{\rm sup}_n \right), \forall k \in \mathcal{K}.
	\end{split}
\end{equation}

By subtracting \eqref{eq:optimize} from \eqref{eq:baseline} and dividing $\delta_{n, i}$ across the stacked equations, we have
{\small 
\begin{equation} 
	\label{eq:zone}
	\begin{split}
		& \frac{T_{n, i}^{\rm set}(k\!+\!1) \!\!-\! T_{n, i}(k)}{\delta_{n, i}}\!\!=\!\! a_{n, ii} \frac{T_{n, i}^{\rm set}\!-\! T_{n, i}(k)}{\delta_{n, i}} \!\!  +\!\!\! \sum_{j \in \mathcal{I}_n(i)}\!\! \frac{\delta_{n, j}}{ \delta_{n, i}}\frac{T_{n, j}^{\rm set} \!-\! T_{n, j}(k)}{\delta_{n, j}}\\
		&\!+\! \frac{b_{n, i}}{\delta_{n, i}}(q_{n, i}(k) - q_{n,i}^{\rm base}(k)), \forall i \in \mathcal{I}_n, k \in \mathcal{K}. 
	\end{split}
\end{equation}}

By substituting  \eqref{eq:soz} into \eqref{eq:zone}, we have 
\begin{equation} \label{eq:zone_soz}
	\begin{split}
		{\rm soz}_{n, i}&(k + 1)\!\! =\!\! a_{n, ii} {\rm soz}_{n, i}(k) \!\! +\!\!\!\!\sum_{j \in \mathcal{I}_n(i)}\!\!\! a_{n, ij} \frac{\delta_{n, j}}{\delta_{n, i}} {\rm soz}_{n, j}(k) \\
		&+ \frac{b_{n, i}}{\delta_{n, i}} (q_{n, i}(k) - q_{n, i}^{\rm base}(k)), ~\forall i \in \mathcal{I}_n, k \in \mathcal{K}. 
	\end{split}
\end{equation}

Finally, by stacking \eqref{eq:zone_soz} across all zones, we  have the following VB models for multi-zone HVAC systems:
\begin{equation} \label{eq:soz_equations}
	\begin{split}
		& {\rm \textbf{Zone~VB~models for multi-zone HVAC system}:}\\
		& \begin{cases}
		 \soz_n(k\!+\!1) \!\!=\!\! \tilde{\A}_n \soz_n(k) \!+\! \tilde{\B}_n\!\!\left(\q_n(k) \!-\! \q_n^{\rm base}(k) \right) \\
		 -\mathbf{1}_n \leq \soz_n(k) \leq \mathbf{1}_n, \\
		 \q_n^{\min}(k) \leq \q_n(k) \leq \q_n^{\max}(k), \quad \forall k \in \mathcal{K}. 
		\end{cases}
	\end{split}
\end{equation}
where we have 
\begin{equation}
	\setlength{\arraycolsep}{2pt} 
	\renewcommand{\arraystretch}{0.9} 
	\begin{aligned}
		\tilde{\A}_n &= \big[a_{n, ij} \delta_{n, j}/\delta_{n, i}\big]_{i,j \in \mathcal{I}_n} \in R^{I_n \times I_n}, \\[2mm]  
		\tilde{\B}_n &= 
		\mathrm{diag}(b_{n,1}/\delta_{n,1}, \dots, b_{n,N_n}/\delta_{n,I_n}) \in R^{I_n \times I_n}.
	\end{aligned}
	\renewcommand{\arraystretch}{1} 
	\setlength{\arraycolsep}{5pt} 
\end{equation}
The accurate upper and low cooling power limits $\q_n^{\min}(k)$ and $\q_n^{\max}(k)$ are usually difficult to obtain as they are temporally coupled. Some simple and close estimations are 
\begin{equation}
	\begin{split}
& \hat{\q}_{n}^{\min}(k)  =  c_p \m_n^{\min}(k)\left(\T_n^{\rm set}(k) - \mathbf{1}_{n} T^{\rm sup}_n \right)   \\
& \hat{\q}_{n}^{\max}(k)  =  c_p \m_n^{\max}(k)\left(\T_n^{\rm set}(k) - \mathbf{1}_{n} T^{\rm sup}_n \right), k \in \mathcal{K}.
	\end{split}
\end{equation}

 
 \begin{remark}
In model \eqref{eq:soz_equations}, the operating flexibility of multi-zone HVAC system has been represented by a series of virtual batteries, corresponding to individual zones. The zone thermal comfort constraints have been encoded by ${\rm soz}_{n, i} \in [-1, 1], \forall i \in \mathcal{I}_n, k\in\mathcal{K}$. This formulation enables flexible multi-zone buildings to function like virtual battery packs. 
 \end{remark} 

\subsection{Aggregated VB Model for Multi-zone HVAC}

To reduce model and computational complexity, we further discuss how the zone VB models can be aggregated into a single  VB at the building level.   To achieve the objective, we first introduce the following important theorem to  be used. 
\begin{theorem} \label{thm:Perron–Frobenius}
	(\textbf{Perron–Frobenius Theorem}) Let $\A \in \R^{n \times n}$ be a non-negative matrix, i.e., $\A_{ij} \geq 0, \forall i, j$. Let $\rho(\A)$ denote the spectral radius of $\A$, defined as
	\begin{equation*}
		\rho(\A) = \max\left\{ \vert \lambda \vert: \lambda~\text{is an eigenvalue of}~\A \right\},
	\end{equation*}
then the following properties hold:  (1) $\rho(\A)$ is an eigenvalue of $\A$; (2) there exists a non-negative eigenvector $\bm{w}$ that $\A \bm{w} = \rho(\A) \bm{w}$.
\end{theorem} 

The parameters  $\tilde{\A}_n$ and  $\tilde{\A}^{\mathsf T}_n$ in \eqref{eq:soz_equations} are  real, non-negative and symmetric matrices, which can be readily verified. Hence, by \textbf{Theorem}~\ref{thm:Perron–Frobenius}, there exist a non-negative eigenvalue $\alpha_n$ and  associated non-negative probability vector $\w_n$ of $\tilde{\A}_n^{\mathsf T}$ such that 
\begin{equation} \label{eq:eigenvector}
	\begin{split}
		& \tilde{\A}^{\mathsf T}_n \w_n = \alpha_n \w_n, \\
		& \mathbf{1}_n^{\mathsf T} \w_n= 1.  
	\end{split}
\end{equation} 
Specifically, the probability eigenvector $\w_n$  can be obtained by normalizing  the  eigenvector of $\tilde{\A}_n^{\mathsf T}$ associated with the  non-negative eigenvalue $\alpha_n$. 
To enable the aggregation of zone VB models, we further propose the following  State-of-Charge  (soc) as the \emph{characterization state} of entire building thermal conditions under zone thermal comfort constraints: 
\begin{equation} \label{eq:soc}
	\begin{split}
		& \soc_n(k) = \w_n^{\mathsf T} \soz_n(k), ~~\forall k \in \mathcal{K}. 
	\end{split}
\end{equation}
\begin{remark}
The soc can be interpreted as the aggregated thermal conditions of entire multi-zone building accounting for zone thermal comfort requirements.  Notably, the physical heterogeneity of thermal zones has been encapsulated by the eigenvector $\w_n$, which enables the proposed modeling framework to be applicable to multi-zone buildings of heterogeneous  zone typologies and envelopes.
\end{remark}

By multiplying both sides of \eqref{eq:soz_equations} with $\w_n^{\mathsf T}$, we have
\begin{equation} \label{eq:soz_equations2}
	    \left\{
\begin{aligned}
		& \w_n^{\mathsf T} \soz_n(k \! + \!1) \!=\! \w_n^{\mathsf T} \tilde{\A}_n \soz_n(k) \!\!+\!\! \w_n^{\mathsf T} \tilde{\B}_n\left(\q_n(k) \!\!-\!\! \q_n^{\rm base}(k) \right) \\
		& \quad - \w_n^{\mathsf T} \bm{1}_{n}\leq \soz_n(k) \leq \w_n^{\mathsf T} \bm{1}_{n}, \\
		& \quad \q_n^{\min}(k) \leq \q_n(k) \leq \q_n^{\max}(k), \quad \forall k \in \mathcal{K}. 
	\end{aligned}
    \right.
\end{equation}

By plugging \eqref{eq:eigenvector}, \eqref{eq:soc} into \eqref{eq:soz_equations2}, we obtain  the following aggregated VB model for  multi-zone HVAC systems: 
\begin{equation}  \label{eq:VB_model_1}
	\left\{  
	\begin{aligned}
		&  \soc_n(k \! + \!1) \!=\! \alpha_n \soc_n(k) \!+\! P_n^{\rm ch/dis}(k) \\
		& \quad  - 1 \leq \soc_n(k) \leq 1, \\
		& \quad \q_n^{\min}(k) \leq \q_n(k) \leq \q_n^{\max}(k), \forall k \in \mathcal{K}.~~~~~~~~~~ 
	\end{aligned}
 \right.
\end{equation}
where $P_n^{\rm ch/dis}(k) \triangleq \w_n^{\mathsf T} \tilde{\B}_n\left(\q_n(k) - \q_n^{\rm base}(k) \right), \forall k \in \mathcal{K}$ can be interpreted as the net charging power of the VB.
One remaining problem is that the net charging power couples  with the individual  zone cooling power $\q_n(k) = [q_{n, i}(k)], \forall i \in \mathcal{I}_n$, which is not favored for building-level energy scheduling. We therefore propose  to approximate the net charging power  by 
\begin{equation} \label{eq:inequality_relaxation}
	\begin{split}
		&  \beta_n^{\min}(k) \mathbf{I}_{n}^{\mathsf T} \q_n(k)\!\!\leq\!\! \w_n^{\mathsf T} \tilde{\B}_n \q_n(k) \leq \beta_n^{\max}(k) \mathbf{I}_{n}^{\mathsf T} \q_n(k),  \\
		&\quad \quad  \q_n^{\min}(k) \leq \q_n(k) \leq \q_n^{\max}(k), \forall k \in \mathcal{K}. 
	\end{split}
\end{equation}
where  $\beta_n^{\min}(k)$ and $\beta_n^{\max}(k)$ are the lower and upper bound parameters to be identified. $\mathbf{I}_{n}^{\mathsf T} \q_n(k)$ represents the total cooling power of all zones.  Many approaches can be used for the parameter estimations.  Some conservative and tight bound parameters  can be obtained as follows. 
\begin{itemize}
\item[1)]  \textbf{Conservative bounds:}  Since $\w_n$ is a nonnegative probability vector and $\tilde{\B}_n$ is a nonnegative diagonal matrix, the product $\w_n^{T}\tilde{\B}_n$ is a non-negative vector. 
 We have the zone cooling power  $\q_n(k)$  nonnegative and bounded, therefore a pair of bound parameters  can be obtained by 
\begin{equation} \label{eq:conservative}
	\begin{split}
	&\beta_n^{\min}(k) = \min_i \left( \mathbf{w}_n^{\mathsf T} \tilde{\B}_n \right)_i,~  \\
	& \beta_n^{\max}(k) =  \max_i \left( \mathbf{w}_n^{\mathsf T} \tilde{\B}_n \right)_i, \quad \forall k \in \mathcal{K}.
	\end{split}
\end{equation}
These bounds are conservative as they hold for all bounded control inputs regardless of the ranges of $\q_n(k)$.

\item[2)]  \textbf{Tight bounds:} The tight and exact bound parameters correspond  to  the following optimization problems. 
\begin{equation} \label{eq:optimal}
	\begin{split}
	&\beta_n^{\min}(k) =\!\!\!\!\!\!\!\!\!\!\! \min_{ \q_n(k) \in [\q_n^{\min}(k), \q_n^{\max}(k)]}\!\!\!\!\!\frac{ \w_n^{\mathsf T} \tilde{\B}_n \q_n(k)}{\mathbf{I}_{n}^{\mathsf T} \q_n(k) } \\ 
	&\beta_n^{\max}(k) = \!\!\!\!\!\!\!\!\!\!\!\max_{ \q_n(k) \in [\q_n^{\min}(k), \q_n^{\max}(k)]}\!\!\!\!\!\frac{ \w_n^{\mathsf T} \tilde{\B}_n \q_n(k)}{\mathbf{I}_{n}^{\mathsf T} \q_n(k) }, \forall k \in \mathcal{K}.
\end{split}
\end{equation}
The tight bounds exist and are available provided with the feasible ranges  of $\q_n(k)$. 	
\end{itemize}

By plugging  \eqref{eq:inequality_relaxation} into \eqref{eq:soz_equations2}, we obtain the  following
aggregated VB   model for multi-zone HVAC systems: 
\begin{equation} \label{eq:VB_model}
	\begin{split}
		& {\rm \textbf{Aggregated VB model for multi-zone HVAC:}} \\
& \begin{cases}
	\soc_n(k+1) = \alpha_n \soc_n(k) + P_n^{\rm ch/dis}(k) \\
	P^{\rm ch/dis}_n(k) \geq \beta_n^{\min}(k) Q_n(k) \!- \!\w_n^{\mathsf T} \tilde{\B}_n\q^{\rm base}_n(k) \\
	P^{\rm ch/dis}_n(k) \leq \beta_n^{\max}(k) Q_n(k) - \w_n^{\mathsf T} \tilde{\B}_n\q^{\rm base}_n(k), \\
	Q_n^{\min}(k) \leq Q_n(k) \leq Q_n^{\max}(k), \\
	-1 \leq \soc_n(k) \leq 1,\quad  \forall k \in \mathcal{K}.
\end{cases}
\end{split}
\end{equation}
where  $Q_n(k) \triangleq \mathbf{I}_{n}^{\mathsf T} \q_n(k)$  denotes the aggregated zone cooling power at time step $k$.  According, we have $Q_n^{\min}(k) =\mathbf{I}_{n}^{\mathsf T} \q_n^{\min}(k)$ and  $Q_n^{\max}(k) =\mathbf{I}_{n}^{\mathsf T} \q_n^{\max}(k)$ represent their lower and upper limits, and some estimations are  $\hat{Q}_n^{\min}(k) =\mathbf{I}_{n}^{\mathsf T} \hat{\q}_n^{\min}(k)$ and  $\hat{Q}_n^{\max}(k) =\mathbf{I}_{n}^{\mathsf T} \hat{\q}_n^{\max}(k)$.

\begin{remark}
In model \eqref{eq:VB_model},  the thermal dynamics, physical operating limits and thermal comfort constraints of multi-zone HVAC operation have been captured  by a virtual battery storage representation,  with $P_n^{\rm ch/dis}(k), \soc_n(k), Q_n(k)$ as  decision variables and the others as static model parameters readily available  from the RC formulations. 
\end{remark}

When considering VB model \eqref{eq:VB_model_1} for building energy management and scheduling,  an energy consumption model  is necessary.  This problem is nontrivial as the  energy consumption of multi-zone HVAC system closely relates to  zone-level control inputs and states (i.e., zone airflow rates, zone temperature), making it difficult to establish an analytic formulation at the aggregated level. 
To overcome the difficulty, this paper adopts a  data-driven approach for energy consumption modeling.  
Specifically, it is clear from   \eqref{eq:cooling_power_fan_power}-\eqref{eq:AHU_power} that  the  system's energy consumption   is determined  by   zone airflow rates $m_{n,i}(k)$, zone temperature $T_{n,i}(k)$, and the outdoor temperature $T^{\rm out}(k)$. The zone-level  control and state variables  in fact have been  encapsulated by the aggregated states $\soc_n(k)$  and aggregated control inputs $Q_n(k)$ with the VB model. Therefore, the energy consumption with  VB model \eqref{eq:VB_model_1} can be captured by the following functional mappings 
\begin{equation} 
	\label{eq:energy_model} 
	\begin{split} 
		Q^{\rm tol}_n(k) \!=\!\mathcal{F}\big( & \soc_n(k\!-\!L:k), Q_n(k\!-\!L:k),  \\ & T^{\rm out}(k\!-\!L: k)\big), \quad \forall k \in \mathcal{K} \end{split} 
\end{equation}
where  $\mathcal{F}(\cdot)$  denotes a class of functional modules, such as neural networks or linear regression models. The hyperparameter $L$ specifies  the look-back window of historical information. We use $k\!-\!L:k$ to denote the time steps spanning from  $k\!-\!L$ to $k$. This paper adopts the long short-term memory (LSTM) networks to capture the functional mappings considering their powerful representation capability. 

%

\begin{remark} \label{rmk:rmk2}
	We have developed  unified VB modeling framework 
	 for characterizing the operating flexibility of both  single-zone and multi-zone building HVAC system, with the thermal comfort constraints encoded by  $\soc \in [-1, 1]$. In fact, the range of soc can be flexibly shifted to our requirements. Specifically, we are able to shift the ranges of soc to  $[0, 1]$  with the VB models  \eqref{eq:AC_VB} and \eqref{eq:VB_model} by
	\begin{itemize}
		\item[1)] Define the  \texttt{characterization state} as 
		\begin{equation} \label{eq:new_soz}
			{\rm soz}_{n}(k) \!\!=\!\! \frac{T_{n}^{\rm max} \!-\! T_{n}(k)}{2\delta_{n}}, \quad \forall  k \in \mathcal{K}.
		\end{equation}
		where $T_{n}^{\rm max} \triangleq  T_{n}^{\rm set} + \delta_{n}, \forall k \in \mathcal{K}$  represent the upper zone comfortable  temperature limits. 
		\item[2)] Set the \texttt{Baseline control} as maintaining the zone temperature at the upper zone temperature limits $T_{n}^{\rm max}$. 
	\end{itemize}
	The above are for single-zone HVAC systems and can be readily extended to multi-zone cases by introducing zone index $i$. 
	Following the same  modeling framework, we can obtain the same VB models \eqref{eq:AC_VB} and \eqref{eq:VB_model}  but with $\soc \in [0, 1]$ and some slight modifications of  model parameters. We refer  readers to  \textbf{Appendix} A and B for details .
\end{remark}


\section{Model Evaluation and Applications}
In this part, we first evaluate the aggregated VB models for characterizing the aggregated thermal dynamics of multi-zone buildings. We then testing the  VB models for DR participation of  building HVAC systems in electricity market.  

\begin{figure}[t]
	\centering
	\includegraphics[width=0.42\textwidth]{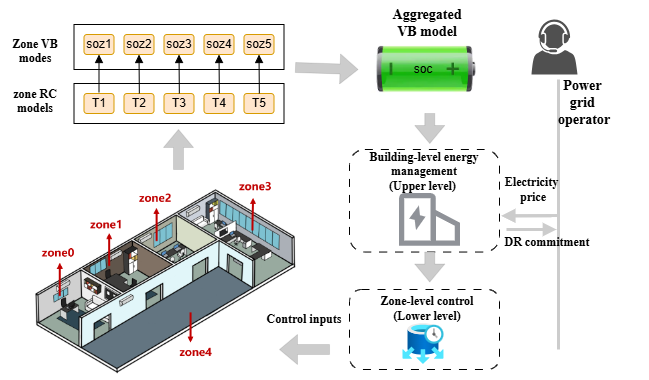}
	\caption{VB model for DR of  multi-zone building HVAC system }
	\label{fig:multi_zone_hvac_5_zones}
\end{figure}

We consider a commercial multi-zone building with  $5$ thermal zones as  illustrated in Fig. \ref{fig:multi_zone_hvac_5_zones}. 
The building thermal parameters and HVAC system configurations are summarized in TABLE \ref{tab:system_configuration}. The thermal disturbances and outdoor weather conditions are taken
from the CityLearn datasets \cite{nweye2025citylearn}.  We set the decision and scheduling  interval to be $\Delta k = 30$ mins with each day equally divided into $K = 48$ time slots. 

\begin{table}[htbp]
	\centering
	\caption{ Building thermal parameters and  HVAC system configurations}
	\label{tab:system_configuration}
	\setlength{\tabcolsep}{4pt}  
	\begin{tabular}{lll}
		\toprule
		Param.  &   Value   & Units \\
		\hline
		$C^{\rm th}_{n, i}~(\forall i \in \mathcal{I}_n)$  & $1.5\times 10^4$  & J/(kg.K) \\
		$c_p$   &  $1.012$   &  kJ/(kg.K)  \\
		$d_r$   & 0.8  & -- \\
		$T_{n, i}^{\rm set}~(\forall i \in \mathcal{I}_n)$  & 25   & $^\circ$C \\
		$\delta_{n, i}~(\forall i \in \mathcal{I}_n)$  & 1 & $^\circ$C  \\ 
		$m_{n, i}^{\min}~(\forall i \in \mathcal{I}_n)$  & 0 & kg/s  \\
		$m_{n, i}^{\max}~(\forall i \in \mathcal{I}_n)$  & 0.5 & kg/s  \\
		$R_{n, ij}^{\rm th}~(\forall i, j \in \mathcal{I}_n)$   &   14   &  kW/K\\
		$R_{n, oi}^{\rm th}~(\forall i \in \mathcal{I}_n)$     &   30   &  kW/K \\ 
		$\kappa_f$   & 0.08 & --\\
		${\rm COP}$   & 1.0  & -- \\
		\bottomrule
	\end{tabular}
\end{table}

\subsection{SOC Model for characterizing building thermal dynamics}

We first evaluate the VB models for characterizing building thermal dynamics under  different control policies of the HVAC system. 
We first obtain the baseline cooling power trajectories $\q_n^{\rm base}(k)$, $\forall k \in \mathcal{K}$  by solving the steady-state equations  \eqref{eq:baseline}  to establish the VB model \eqref{eq:VB_model}. 
 We consider  three representative control policies for the  multi-zone HVAC system: 
 
 \begin{itemize}
\item \texttt{Random control}: Randomly generating  zone airflow rates   within the operating ranges of VAV boxes for  multi-zone HVAC system.    
\item \texttt{PID control}: Tuning a PID controller to modulate the zone airflow rates to keep the zone temperatures close to their respective comfort ranges.  
\item \texttt{DRL control}: Training a deep reinforcement learning (DRL) agent to control zone airflow rates  with the objective minimizing the weighted sum of total electricity cost and comfortable zone temperature violations. 
\end{itemize}
 
We simulate  the three control policies on the RC model \eqref{eq:optimize} to obtain the  
zone temperature $T_{n, i}(k)$ and cooling power trajectories $q_{n, i}(k)$. 
Each simulation spans $5$ days ($K = 240$ time steps) to  account for the uncertainties of thermal disturbances and weather conditions.
Based on the simulated  trajectories,  we identify the actual soc trajectories  ${\rm soc}^{\rm true}$ for the building according to \eqref{eq:soz} and \eqref{eq:soc}, which present the true aggregated thermal dynamics of the building under the control policy.  We then compare the true  ${\rm soc}^{\rm true}$  with the soc obtained from the VB models for model evaluation.

We  first establish the mappings from the control inputs of RC model to  the VB model \eqref{eq:VB_model} to keep  consistent operation. Specifically, for  each  simulated trajectory of the RC model,  we have $Q_n(k) = \mathbf{1}^\top \q_n(k)$ for the VB model.  We consider  three algorithms for estimating the model parameters  $\beta_n^{\min}(k)$ and $\beta_n^{\max}(k)$:  
\begin{itemize}
	\item \textbf{Algorithm~1 (Conservative method):} Calculating $\beta_n^{\min}(k)$ and $\beta_n^{\max}(k), \forall k \in \mathcal{K}$ with the conservative approach introduced in \eqref{eq:conservative}.
	\item \textbf{Algorithm~2 (Step-ahead estimation):}  Estimating $\beta_n^{\min}(k)$ and $\beta_n^{\max}(k), \forall k \in \mathcal{K}$ using \eqref{eq:optimal} by fixing  $\q_n(k)$ to the  zone cooling power $\q_n(k-1)$ obtained from previous time step. 

	\item \textbf{Algorithm~3 (Optimal tight bounds)} Computing $\beta_n^{\min}(k)$ and $\beta_n^{\max}(k), \forall k \in \mathcal{K}$  according to \eqref{eq:optimal} by setting  $\q_n(k)$ to be the actual zone cooling power obtained with the RC model.
\end{itemize}

\begin{figure*}[t]
	\centering
	\begin{subfigure}[b]{0.32\textwidth}
		\centering
		\includegraphics[width=\textwidth]{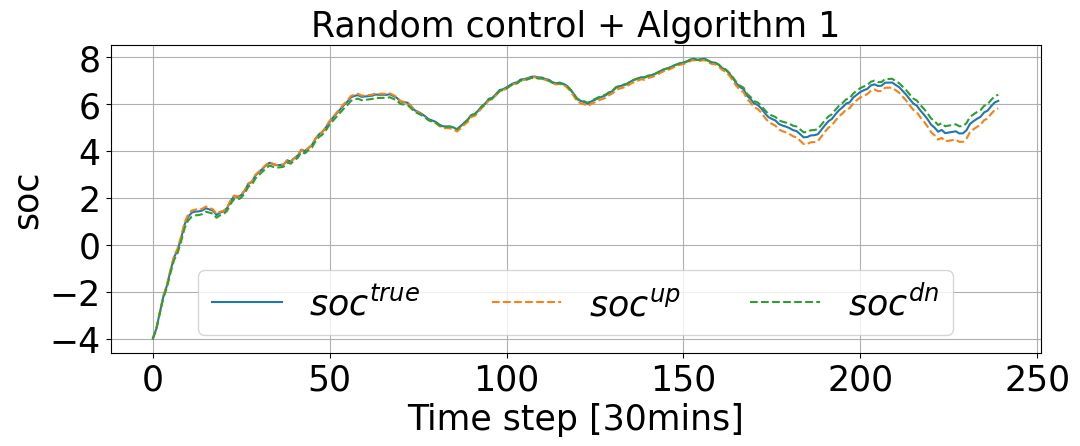}
		\label{fig:random1}
	\end{subfigure}
	\begin{subfigure}[b]{0.32\textwidth}
		\centering
		\includegraphics[width=\textwidth]{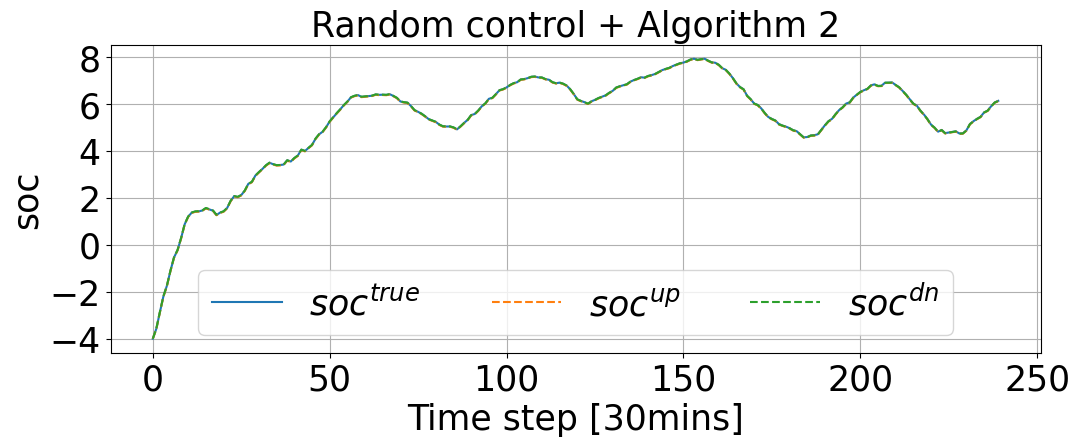}
		\label{fig:random2}
	\end{subfigure}
	\begin{subfigure}[b]{0.32\textwidth}
		\centering
		\includegraphics[width=\textwidth]{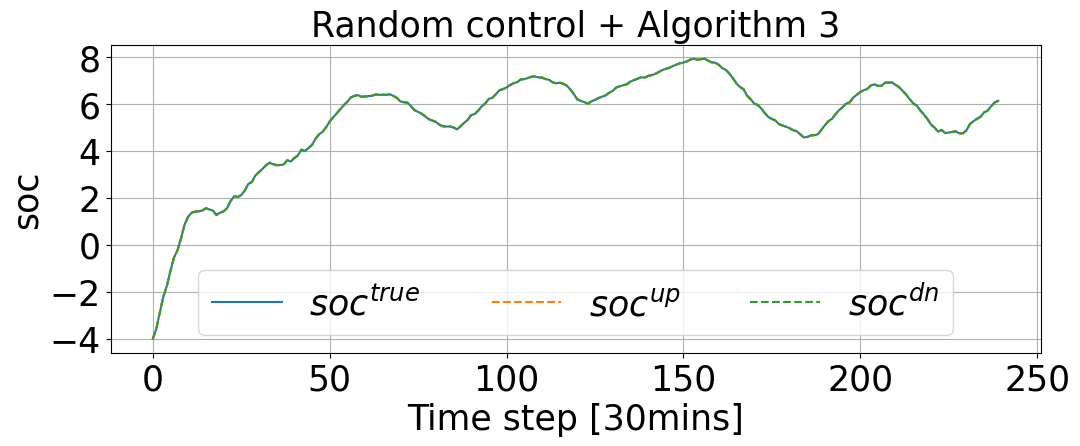}
		\label{fig:random3}
	\end{subfigure}

	\begin{subfigure}[b]{0.32\textwidth}
		\centering
		\includegraphics[width=\textwidth]{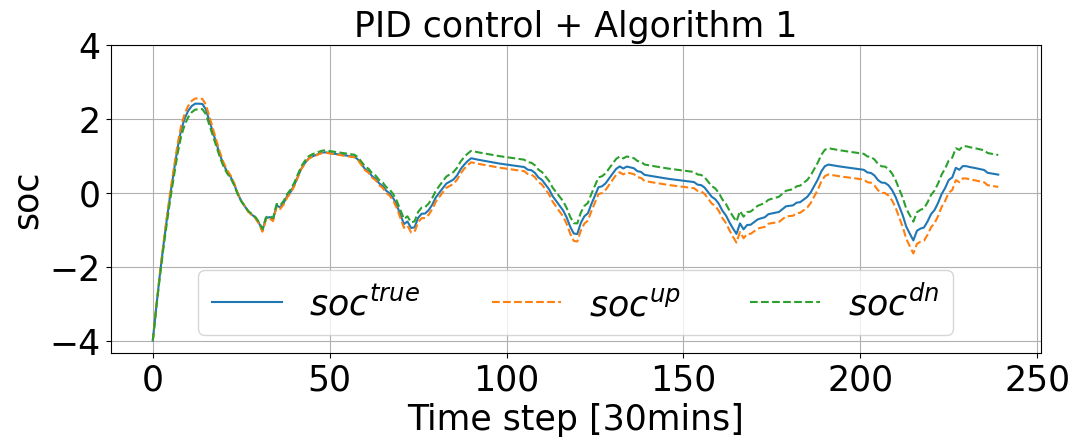}
		\label{fig:pid1}
	\end{subfigure}
	\begin{subfigure}[b]{0.32\textwidth}
		\centering
		\includegraphics[width=\textwidth]{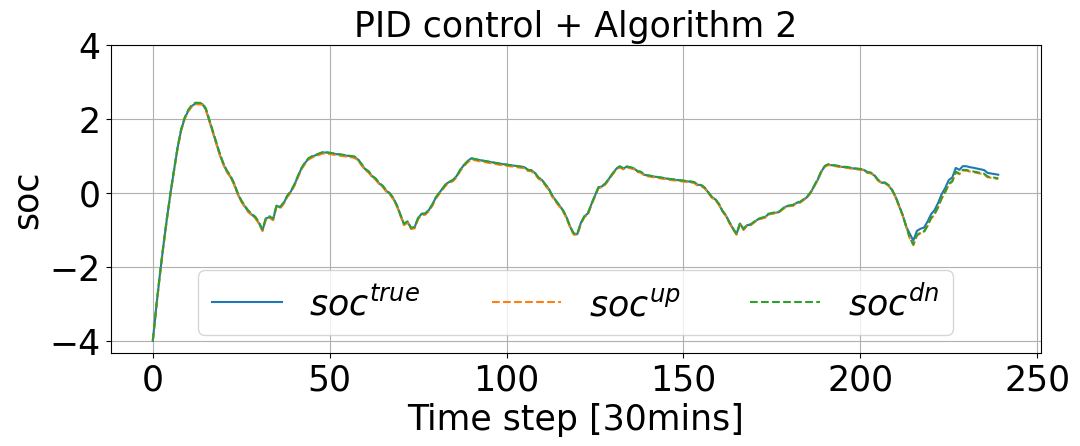}
		\label{fig:pid2}
	\end{subfigure}
	\begin{subfigure}[b]{0.32\textwidth}
		\centering
		\includegraphics[width=\textwidth]{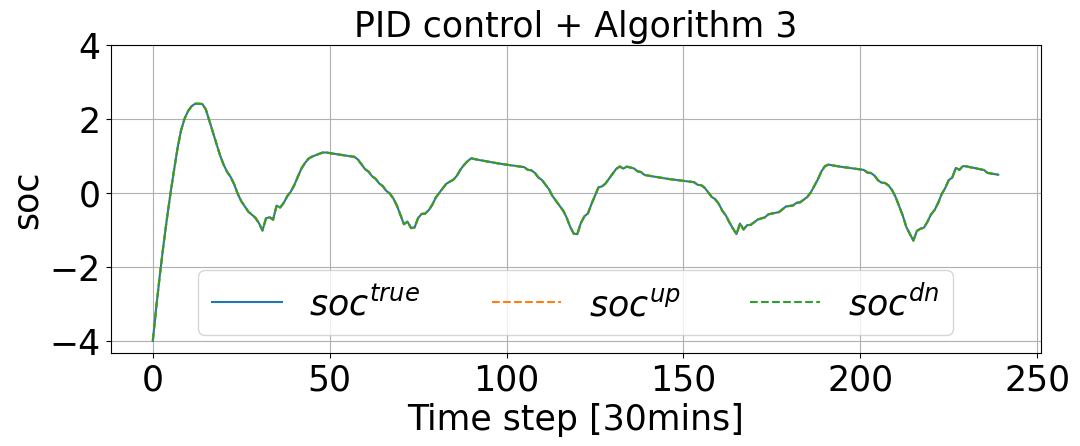}
		\label{fig:pid3}
	\end{subfigure}

	\begin{subfigure}[b]{0.32\textwidth}
		\centering
		\includegraphics[width=\textwidth]{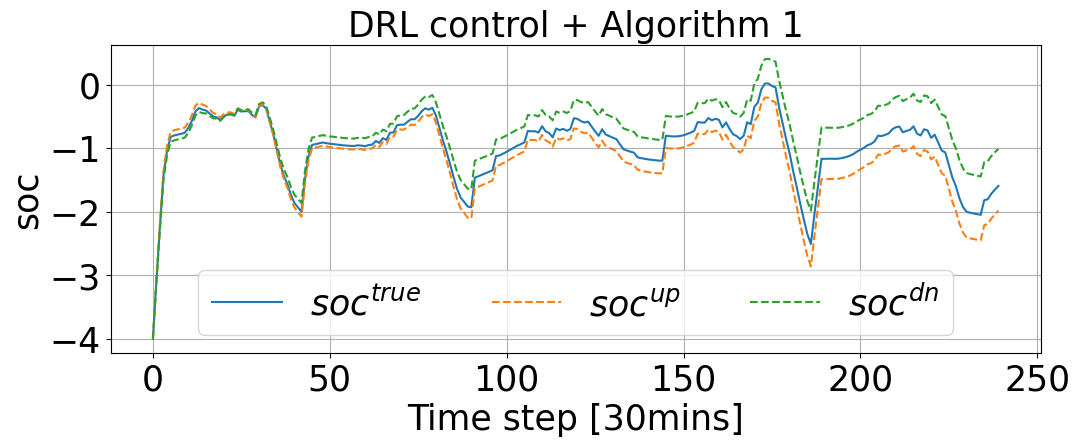}
		\label{fig:drl1}
	\end{subfigure}
	\begin{subfigure}[b]{0.32\textwidth}
		\centering
		\includegraphics[width=\textwidth]{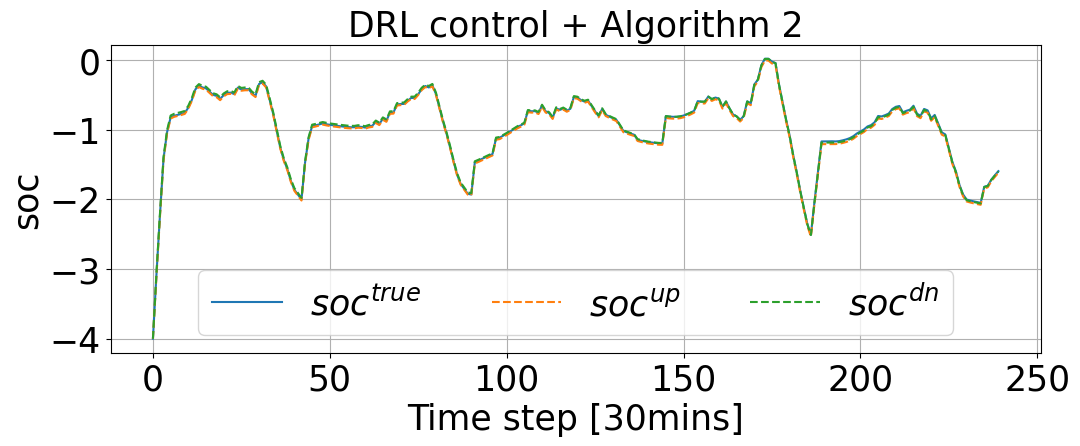}
		\label{fig:drl2}
	\end{subfigure}
	\hfill
	\begin{subfigure}[b]{0.32\textwidth}
		\centering
		\includegraphics[width=\textwidth]{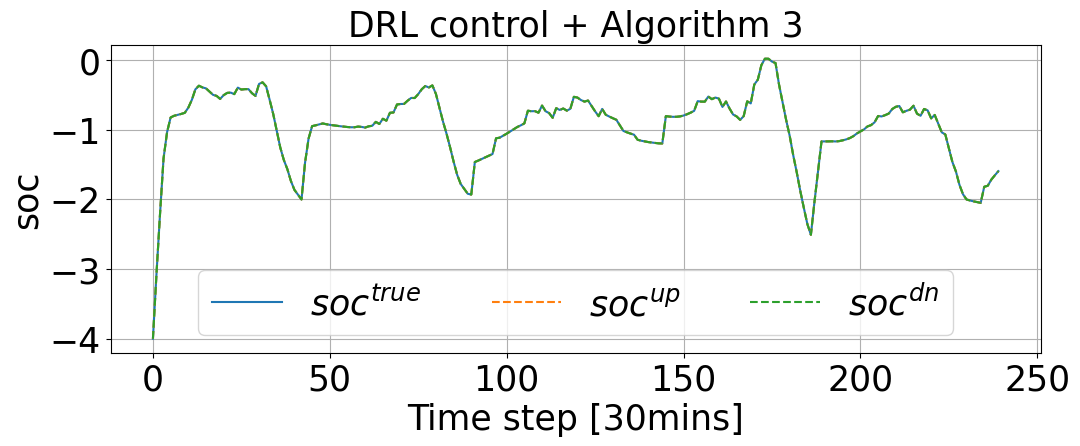}
		\label{fig:drl3}
	\end{subfigure}
	
	\caption{The actual soc  from the RC model and the upper and low soc from the VB model under different control policies (\texttt{Random control}, \texttt{PID control}, \texttt{DRL control}) and methods (\texttt{Algorithm 1}, \texttt{Algorithm 2}, \texttt{Algorithm 3}) for estimating model parameters   $\beta_n^{\min}(k)$ and $\beta_n^{\max}(k)$.  }
	\label{fig:soc_estimation}
\end{figure*}

With the control inputs  $Q_n(k), \forall k \in \mathcal{K}$ and the VB model parameters, we are able to simulate the soc trajectory of the multi-zone building with the  VB model.  However, one remaining issue is that,  for the VB model \eqref{eq:VB_model}, the $\soc$ is not uniquely  determined by the  control inputs $Q_n(k)$ due to the inequality relaxation \eqref{eq:inequality_relaxation}. To handle this issue, we choose to evaluate the  upper and lower bounds of  $\soc$ for the given control inputs $Q_n(k)$: 
\begin{equation*} 
	\begin{split}
		& \soc^{\rm up}(k + 1) \!=\! \soc^{\rm up}(k) \!+\! P_n^{\rm ch/dis, up}(k), \\
		& \soc^{\rm dn}(k + 1) \!=\! \soc^{\rm dn}(k) \!+\! P_n^{\rm ch/dis, dn}(k), \forall k \in \mathcal{K}.~~~~~~~~~ \\
	\end{split}
\end{equation*}
where we have 
\begin{equation*}
	\begin{split}
		&  P^{\rm ch/dis, up}_n(k) \!=\! \beta_n^{\max}(k) Q_n(k) \!-\! \w_n^{\mathsf T} \tilde{\B}_n\q^{\rm base}_n(k),\\
		&  P^{\rm ch/dis, dn}_n(k)\!=\! \beta_n^{\min}(k) Q_n(k) \!-\! \w_n^{\mathsf T} \tilde{\B}_n\q^{\rm base}_n(k), \forall k \in \mathcal{K}.\\
	\end{split}
\end{equation*}
Note that the differences between $\soc^{\rm up}(k)$ and $\soc^{\rm dn}(k)$ directly  relate to the model parameters $\beta_n^{\min}(k)$ and $\beta_n^{\max}(k)$. We  compare $\soc^{\rm up}$ and $\soc^{\rm dn}$ obtained from the VB models with  ${\rm soc}^{\rm true}$ to evaluate the the capability of VB models for capturing building thermal dynamics. We present the results for each control policy and the algorithm for model parameter estimations  in Fig.\ref{fig:soc_estimation}.    It can be observed that the  $\soc^{\rm up}$ and $\soc^{\rm dn}$ obtained from the VB models closely match the actual trajectories $\soc^{\rm true}$ for almost all cases, demonstrating that the  VB models can well capture the building thermal dynamics. Some gaps  occur with the results of \texttt{Algorithm  1} (see Fig. \ref{fig:soc_estimation} (a), (d), (g)), which are due to the  quite  conservative model parameters $\beta_n^{\min}(k)$ and  $\beta_n^{\max}(k)$ provided by the method.     
These  gaps   can be effectively decreased by employing tighter model parameters $\beta_n^{\min}(k)$ and  $\beta_n^{\max}(k)$ as can be seen from  the results with  \texttt{Algorithm 2} (see Fig. \ref{fig:soc_estimation} (b), (e), (h)), in which the step-ahead estimation method is used and gives  tighter model parameters.  This conclusion has been further confirmed by 
the results of \texttt{Algorithm 3}  (see Fig. \ref{fig:soc_estimation} (e), (f), (i)), in which the optimal and tight model parameters are employed. As a result,  
 the upper and lower soc trajectories  $\soc^{\rm up}$ and $\soc^{\rm dn}$ exactly coincide with the actual $\soc^{\rm true}$. In such case, the proposed VB model is  equivalent to the original RC model in characterizing building thermal dynamics.

The above results demonstrate that the VB models can well characterize the   building thermal dynamics with  sufficiently tight  model parameters. Conservative model parameters may lead to  certain accumulated gap in terms of soc trajectories, whereas it seems still applicable to many building-level energy management tasks 
as the accumulated gap is still quite minor for  sufficiently long   periods (see the previous 100 time steps of Fig. \ref{fig:soc_estimation} (a), (d), (g)).

\subsection{Energy Consumption Model for Multi-zone HVAC}
We further evaluate the data-driven approach for  energy consumption modeling with the VB model. We use a LSTM network  (\texttt{layers: 1 seq length: 2, hidden dim: 64, batch size: 64}) combined with a fully connected layer to capture the functional  mappings \eqref{eq:energy_model}. We set the look-back window  $L=1$  as we observe that one-step historical information leads to the best model performance in the experiments. 
  We first build datasets for model training,  and then evaluate the model accuracy and generality.  
Specifically, we run the three control policies (i.e.,\texttt{Random control}, \texttt{PID control} and \texttt{DRL control}) using the RC model each spanning $100$ days for data collection.  For each simulated trajectory,  the following state and control variables are evaluated and collected as samples: 
\begin{equation}
	\big(\soc_n(k), Q_n(k), T^{\rm out}(k),  Q^{\rm tol}_n(k)\big), \quad \forall k \in \mathcal{K}.
\end{equation}
Based on the collected samples, we generate four datasets. Three datasets correspond directly to the trajectories produced by the three control policies and denoted as \texttt{Random control}, \texttt{PID control}, and \texttt{DRL control}, respectively. The fourth dataset, referred to as \texttt{Mixture control}, is obtained by uniformly concatenating segments from the three individual datasets. Note that the four datasets are of equal length, each of which contains $4.8$K samples. We further divide each dataset by the ratio of $6:2:4$ for model training, validating and testing.

We then train four energy consumption models with the four datasets using  the same LSTM architecture, and  subsequently evaluate them on the same \texttt{Mixture Control} dataset. 
Particularly, we do not make any normalization on the datasets. 
Multiple widely-used metrics are employed to quantify the model performance  and the results are reported in TABLE~\ref{tab:LSTM_model_accuracy}. The best case in terms of each performance metric has been highlighted  in bold text.  
 First of all, we see that all the energy consumption models learned from the four datasets show favorable performance. The Pearson correlation coefficient (\texttt{Corr}) is quite close to 1. Meanwhile, the mean absolute percentage error (\texttt{MAPE}) are all less than 10\%.  
 By inspecting the performance metrics further, we observe that  \texttt{Mixture control} dataset provides  slightly better performance than the other three. This is  reasonable  as the \texttt{Mixture control} dataset is generated from three control policies and shows larger diversity in terms of system operation patterns.
 
\begin{figure*}[t]
	\centering
	\begin{subfigure}[b]{0.48\textwidth}
		\centering
		\includegraphics[width=\textwidth]{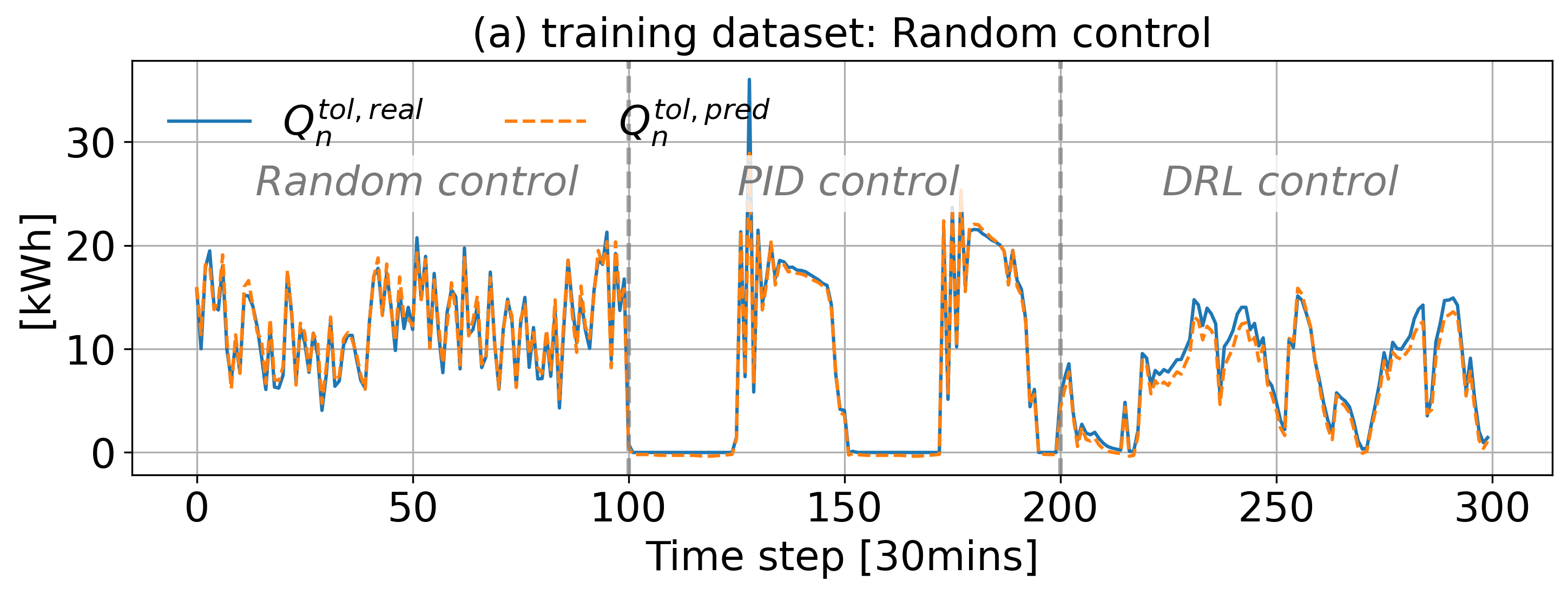}
		\label{fig:lstm_random}
	\end{subfigure}
	\hfill
	\begin{subfigure}[b]{0.48\textwidth}
		\centering
		\includegraphics[width=\textwidth]{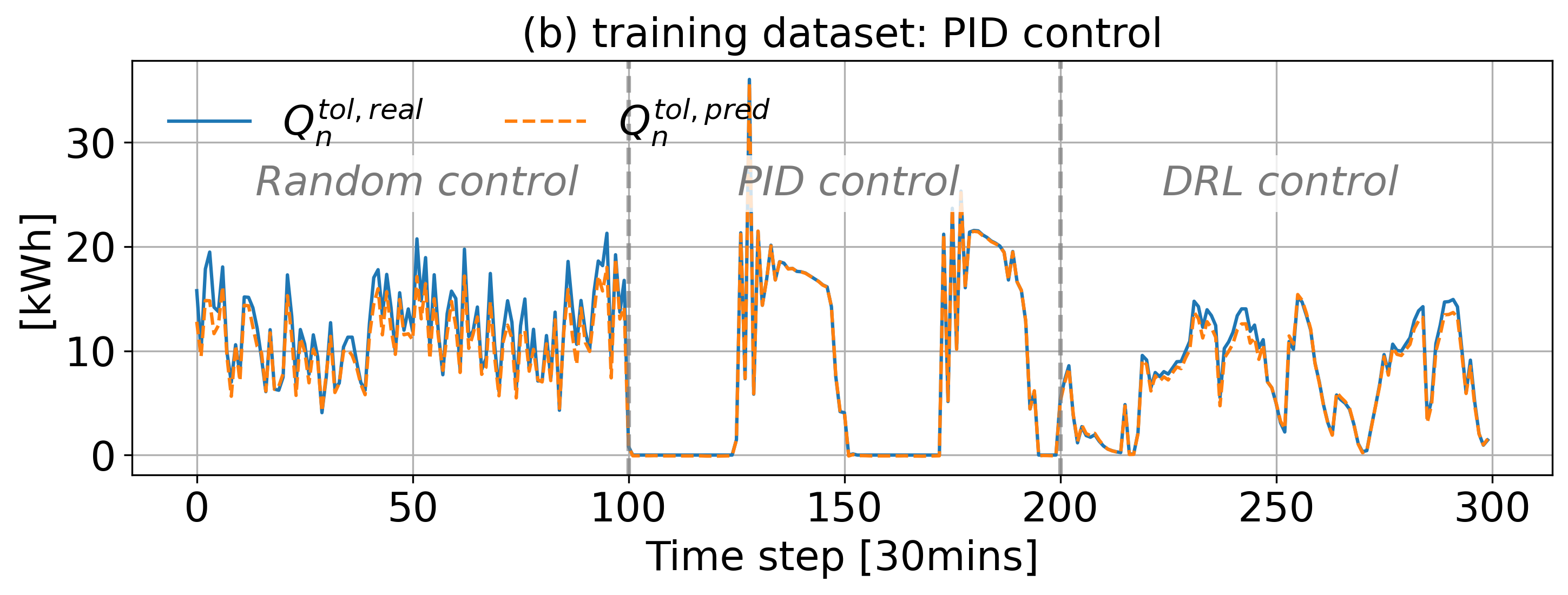}
		\label{fig:lstm_pid}
	\end{subfigure}
	
	\begin{subfigure}[b]{0.48\textwidth}
		\centering
		\includegraphics[width=\textwidth]{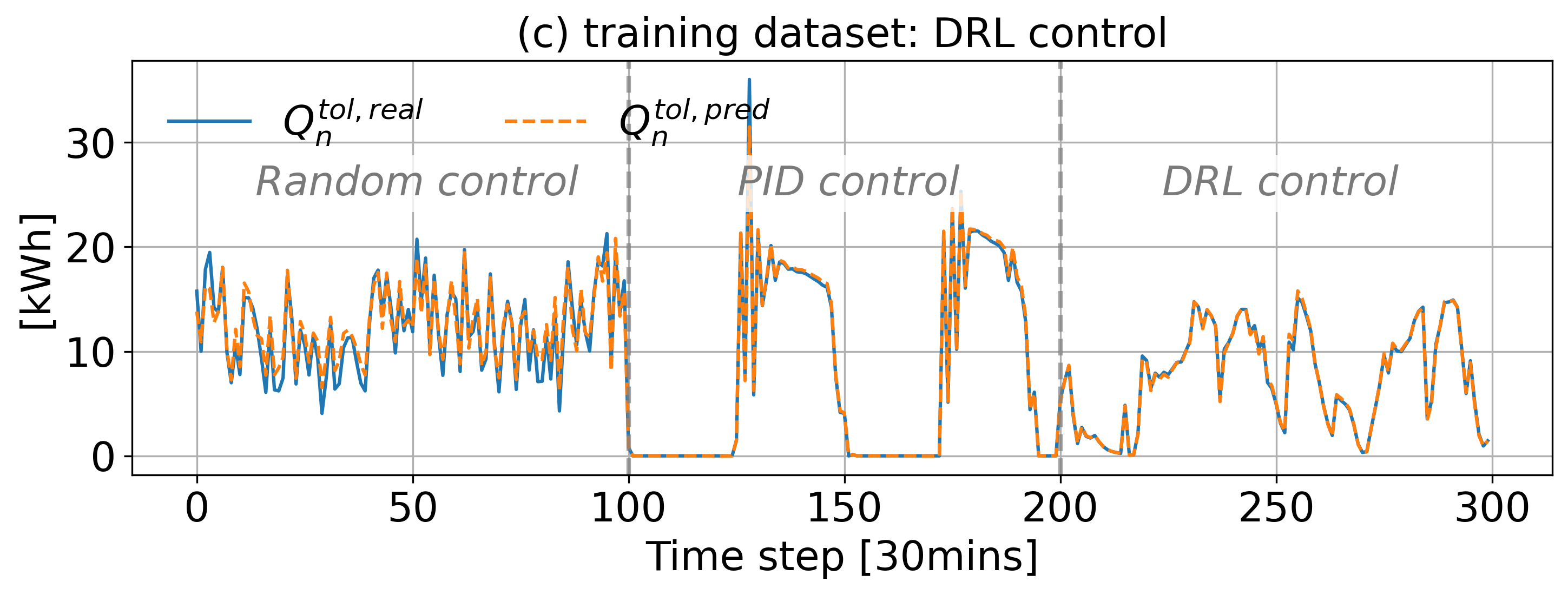}
		\label{fig:lstm_DDPG}
	\end{subfigure}
	\begin{subfigure}[b]{0.48\textwidth}
		\centering
		\includegraphics[width=\textwidth]{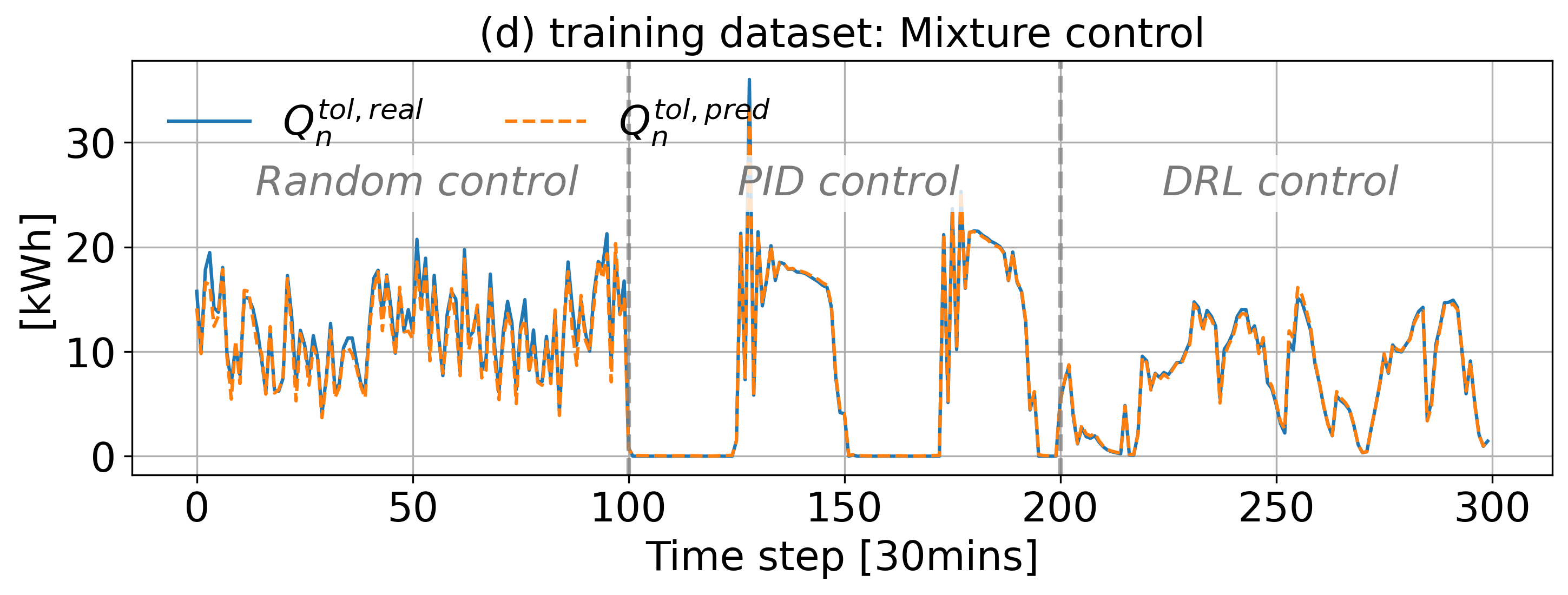}
		\label{fig:lstm_mixed}
	\end{subfigure}
	
	\caption{Predictive performance of energy consumption models trained from different datasets: (a) \texttt{Random control}, (b) \texttt{PID control}, (c) \texttt{DRL control} testing on the \texttt{Mixture control} dataset.}
	\label{fig:lstm_accuracy}
\end{figure*}

We further evaluate the predictive performance of the trained models. We 
use the trained  models to infer the energy consumption under different control policies, and compare them with the actual values.   The results are presented in Fig. \ref{fig:lstm_accuracy}. Overall, all trained models demonstrate satisfactory predictive capability, and  the model trained from the \texttt{Mixture control} dataset behaves the best, which is consistent with the performance metrics presented in TABLE~\ref{tab:LSTM_model_accuracy}. Besides, it is important to note that all the trained  models demonstrate satisfactory generality. 
Specifically, the  models trained from the datasets generated from a specific control policy perform quite well in capturing the energy consumption of other control polices. This demonstrates the effectiveness of the data-driven approach in modeling the energy consumption of multi-zone HVAC systems with the VB model.

\begin{table}[htbp]
	\centering
	\caption{Performance of trained energy consumption models for multi-zone HVAC system}
	\label{tab:LSTM_model_accuracy}
	
	\setlength{\tabcolsep}{4pt}  
	
	\begin{tabular}{lcccccc}
		\toprule
		\multirow{2}{*}{Datasets (Training)} & \multicolumn{6}{c}{\textbf{Mixture control (Testing)}} \\
		\cline{2-7}
		& \makecell[c]{MAPE\\{[\%]}}  & \makecell[c]{RMSE\\{[kW]}} &  \makecell[c]{MAE\\{[kW]}} & \makecell[c]{RSE\\{[\%]}} & \makecell[c]{RAE\\{[\%]}} & \makecell[c]{Corr\\{[0, 1]}} \\
		\hline
		\textbf{Random control} & 7.89 & \textbf{0.6606} & 0.5101 & \textbf{10.37} & 9.49 & \textbf{0.9954} \\
		\textbf{PID control}    & 5.36 & 1.0162 & 0.5704 & 15.95 & 10.61 & 0.9918 \\
		\textbf{DRL control}    & 7.75 & 1.5334 & 0.8644 & 24.06 & 16.08 & 0.9794 \\
		\textbf{Mixture control}& \textbf{4.51} & 0.7358 & \textbf{0.4294} & 11.55 & \textbf{7.99} & 0.9944 \\
		\bottomrule
	\end{tabular}
\end{table}

\subsection{VB model for Building Demand Response (DR)}
In this part, we apply the  VB model for  DR participation  of  the multi-zone building in electricity market. Specifically, we consider  Time-of-Use (ToU)  electricity price and study how the control flexibility of the multi-zone HVAC system can be exploited to reduce the building's electricity cost. This problem can  be modeled as a hierarchical control problem: the \texttt{upper level} determines the DR commitment of the  multi-zone building HVAC system and the \texttt{lower level} divides the DR commitment to  zone-level control inputs when  called by the power grid operator. 
The \texttt{upper level} DR commitment can be  made by using the proposed VB model with the following problem formulation:  
\begin{equation} \label{eq:upper_level}
	\begin{split}
		& \hat{\mathbf{X}} \!\!=\!\! \arg \min_{\mathbf{X}}\sum_{k \in \mathcal{K}} c(k) \cdot Q^{\rm tol}_n(k) \quad {\rm s.t.}~ \eqref{eq:VB_model}-\eqref{eq:energy_model}. \\
	\end{split}
\end{equation}
where  $\mathbf{X} \triangleq [ \soc_n(k),  P_n^{\rm ch/dis}(k), Q_n(k), Q^{\rm tol}_n(k)], \forall k \in \mathcal{K}$ are decision variables. $c(k), \forall k \in \mathcal{K}$ denote the ToU electricity price. The model parameters $\beta_n^{\min}(k)$ and $\beta_n^{\max}(k)$ are computed using the conservative method of \texttt{Algorithm 1}. The resulted DR commitment is denoted by $[\hat{Q}^{\rm tol}_n(k)], \forall k \in \mathcal{K}$, which represents the committed energy consumption trajectory of the HVAC system over the DR window $\mathcal{K}$.

When the committed DR is called by power grid operator, the \texttt{lower level} requires to decompose the DR strategy to zone-level control inputs.
This can be formulated as  tracking the committed energy consumption trajectory $\hat{Q}^{\rm tol}_n(k)$  as close as possible while  respecting the operating limits and thermal comfort constraints: 
\begin{equation} \label{eq:lower_level}
	\begin{split}
		& \hat{\mathbf{Y}}^{*} = \arg \min_{\mathbf{Y}}~ \sum_{k \in \mathcal{K}} \left(Q^{\rm tol}_n(k) - \hat{Q}^{\rm tol}_n(k)\right)^2\\
		& {\rm s.t.}~   \eqref{eq:zone_dynamics}-\eqref{con:zone_comfort}, \eqref{eq:cooling_power_fan_power}-\eqref{eq:AHU_power}. \\
	\end{split}
\end{equation}
where $\mathbf{Y} \triangleq [m_{n, i}(k), T_{n, i}(k), q_{n, i}(k), Q^{\rm tol}_n(k)], \forall i \in \mathcal{I}_n, k \in \mathcal{K}$ denote the decision variables. The \texttt{lower level} uses  RC model to obtain  zone-level control inputs. We denote the resulted actual energy consumption trajectory of  HVAC system as $\hat{\mathbf{Q}}^{\rm tol,*}_n(k), \forall k \in \mathcal{K}$. 
The actual electricity cost for DR participation  is evaluated as ${\rm Cost}_{\rm VB} = \sum_{k \in \mathcal{K}} c(k)\cdot \hat{\mathbf{Q}}^{\rm tol,*}_n(k)$.

We evaluate the effectiveness of the VB model for DR participation  by comparing the electricity cost ${\rm Cost}_{\rm VB}$ with the theoretical optima ${\rm Cost}_{\rm Opt}$, which can obtained  by solving the optimization problem \eqref{eq:lower_level} with the objective replaced by that of \eqref{eq:upper_level}. 
We consider electricity markets with different DR commitment windows, including 1 day, 3 days and 5 days (i.e., $K = 48, 144, 240$). For each case, $S = 30$ independent experiments are performed to account for the uncertainties.  
We calculate the average electricity cost of $S= 30$ experiments and use it as  performance metrics. In addition to electricity cost, we also evaluate the model complexity  by the number of decision variables involved. Particularly, we  only focus on DR commitments and thus only evaluate the model complexity  of \texttt{upper level} for the VB model.  
The results of all case studies are summarized in TABLE~\ref{tab:DR_performance_gap}. 
We note that there only exist  minor performance gaps (about 5\%) in terms of the average electricity cost with the VB model compared with the theoretical optima. Moreover, we see that the performance of  VB model for DR commitment is quite stable under different DR windows.  The results demonstrate that the conservative model parameters $\beta_n^{\min}(k)$  and $\beta_n^{\min}(k)$  are sufficient for effective DR participation of buildings. 
Besides, it is important to note that the  VB model  can enable  considerable model complexity reduction compared with RC model. 
Specifically, the number of decision variables with the VB model only linearly increase with the optimization horizon  regardless the number of thermal zones involved, whereas the number of decision variables increases both with the number of zones and the optimization horizon with the RC model.  This is because the zone operating flexibility has been aggregated by the VB model. 
Moreover, the VB model 
  admits  a linear state-space model, which is in contrast to the nonlinear state-space model with the RC model. Therefore, the  VB model favors  building-level energy management and scheduling considering the model and computational complexity.  
  
\begin{table}[htbp]
	\centering
	\caption{Performance of DR for multi-zone HVAC systems under different DR commitment windows} 
	\label{tab:DR_performance_gap}
		\setlength{\tabcolsep}{2pt}  
		
	\begin{tabular}{rrrrrrr}
		\toprule	
		 &  \multicolumn{2}{c}{\textbf{RC model}} & \multicolumn{2}{c}{\textbf{VB model}} &   \\[3pt]
		\cline{2-5}
		\makecell[c]{Optimization \\ Horizon [Steps]} 
		&  \makecell[c]{${\rm Cost}_{\rm Opt}$ \\ (Avg.) [\$]} 
		& \makecell[c]{\# Decision \\ variable} 
		& \makecell[c]{${\rm Cost}_{\rm VB}$ \\ (Avg.) [\$]} 
		& \makecell[c]{\# Decision \\ variable}  
		& \makecell[c]{ Performance \\ Gap [\%]} \\ 
		\midrule
		~48 (1~day)     & 57.71  &  480  & 61.08  & 144   & 5.7\%  \\
		144 (3 days)    & 194.14  &  1440  & 201.14  & 432   & 3.7\%  \\
		240 (5 days)    & 328.39  &  2400  & 339.44  & 720   & 3.4\%  \\
		\bottomrule
	\end{tabular}
\end{table}

\begin{figure}[htbp]
	\centering
	\begin{subfigure}[b]{0.48\textwidth}
		\includegraphics[width=\textwidth]{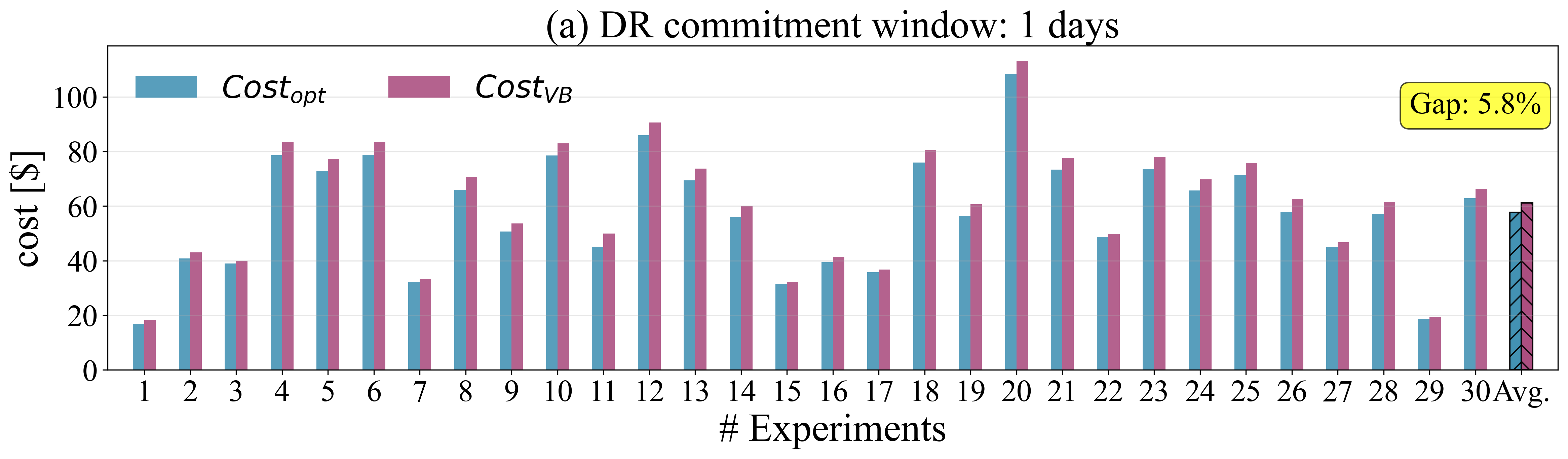}
		\label{fig:optimal_vs_tracking_1days}
	\end{subfigure}
	\hfill
	\begin{subfigure}[b]{0.48\textwidth}
		\includegraphics[width=\textwidth]{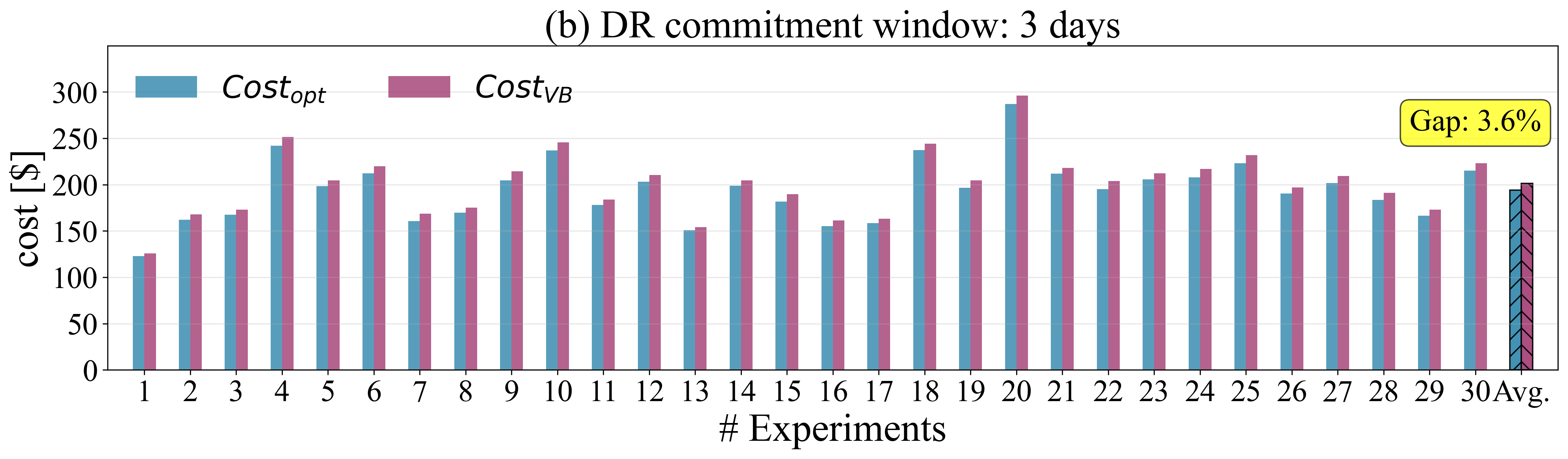}
		\label{fig:optimal_vs_tracking_3days}
	\end{subfigure}
	\hfill
	\begin{subfigure}[b]{0.48\textwidth}
		\includegraphics[width=\textwidth]{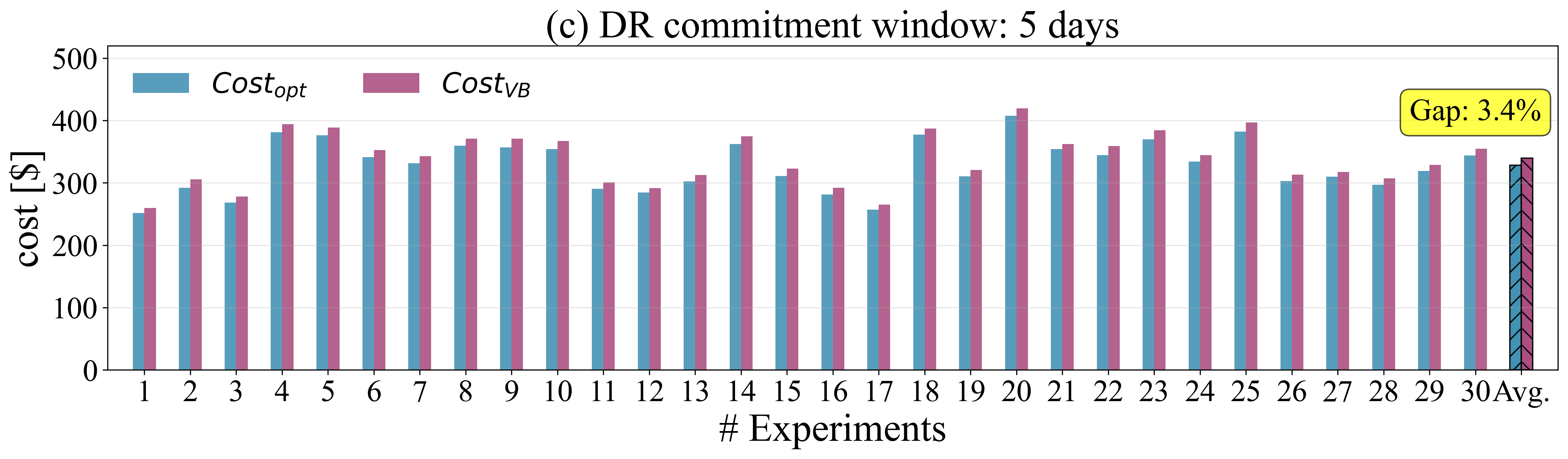}
		\label{fig:optimal_vs_tracking_5days}
	\end{subfigure}
	\caption{Electricity cost distributions of  VB model and theoretical optima in electricity markets with different DR commitment windows: (a) $K =48$ (1 day),  (b) $K = 144$ (3 days),  (c) $K = 240$ (5 days). }
	\label{fig:Q_total_op_vs_track}
\end{figure}

To further demonstrate the favorable performance of  VB models for DR participation, we 
compare the electricity cost distribution of the $S=30$ experiments obtained from the VB model with  the theoretical optima.  The  results are presented in   Fig.~\ref{fig:Q_total_op_vs_track}. We see that for the different DR commitment windows, the electricity cost distributions  with the  VB model are quite  close to  the theoretical optima. This demonstrates that the proposed VB model can provide near-optimal and stable performance for DR participation under uncertainties and varying DR commitment windows. 

\begin{figure}[htbp]
	\centering
	\begin{subfigure}[b]{0.48\textwidth}
		\includegraphics[width=\textwidth]{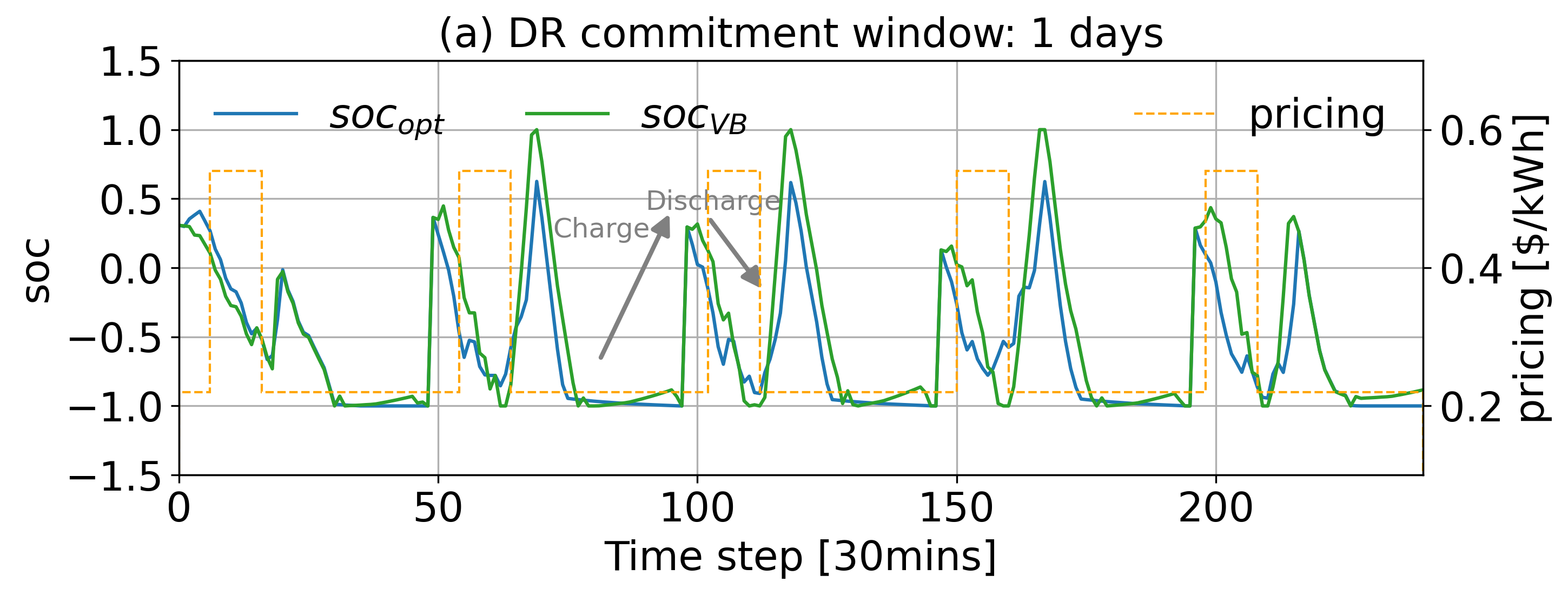}
		\label{fig:soc_pricing_VB_1day}
	\end{subfigure}
	\begin{subfigure}[b]{0.48\textwidth}
		\includegraphics[width=\textwidth]{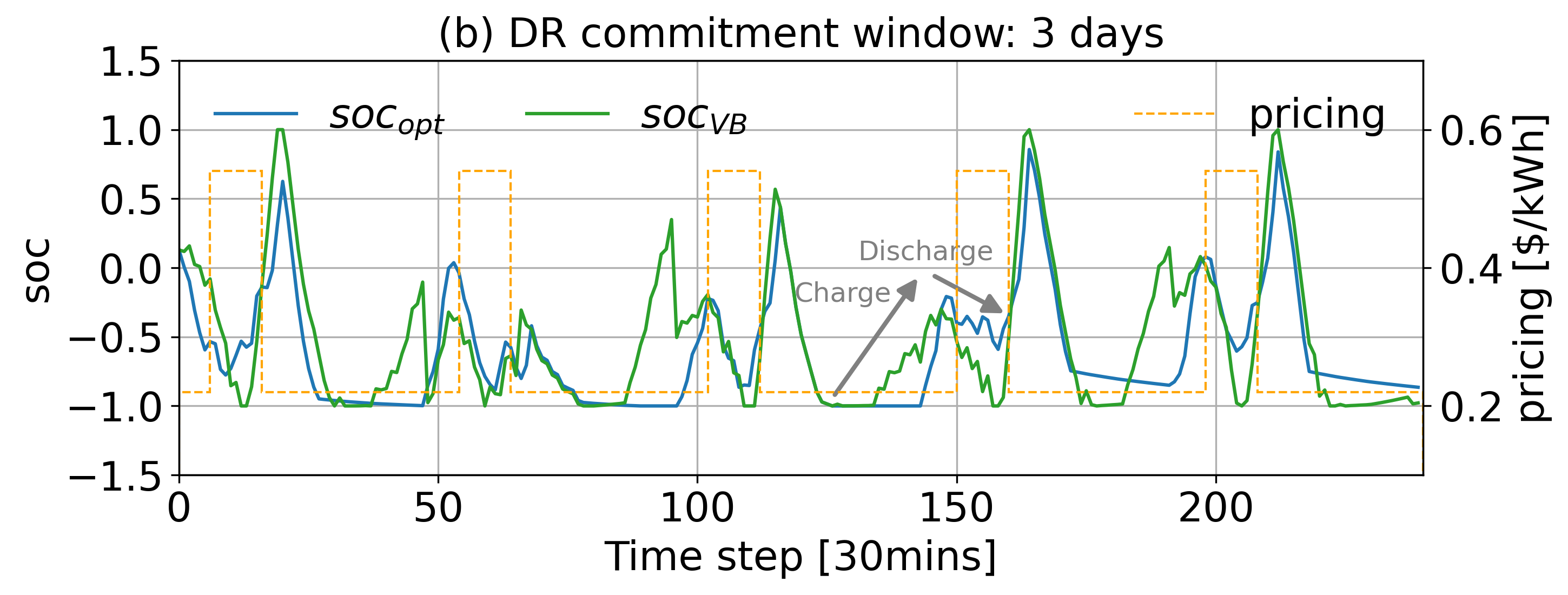}
		\label{fig:soc_pricing_VB_3day}
	\end{subfigure}
	\begin{subfigure}[b]{0.48\textwidth}
		\includegraphics[width=\textwidth]{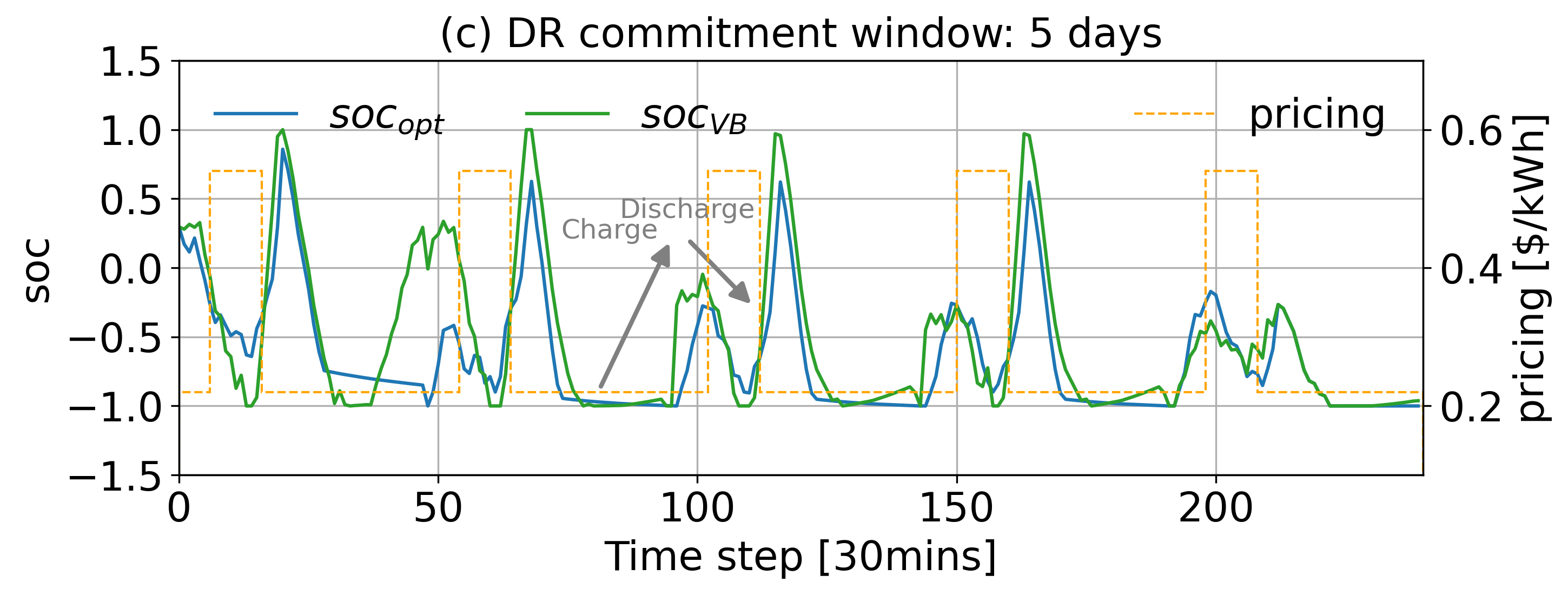}
		\label{fig:soc_pricing_VB_5day}
	\end{subfigure}
	\caption{ The dynamic soc of multi-zone HVAC system  in response to DR price in electricity markets with different DR commitment windows: (a) $K =48$ (1 day),  (b) $K = 144$ (3 days),  (c) $K = 240$ (5 days). }
	\label{fig:soc_for_DR}
\end{figure}

Another novel benefit of the VB model is that it enables flexible buildings to function like virtual batteries. To examine this behavior, we investigate  the soc of multi-zone HVAC system under varying DR prices. We randomly select five days from the $S = 30$ experiments of different DR commitment windows and present the  building soc trajectories  in Fig. \ref{fig:soc_for_DR}. We can observe an obvious inverse patterns in terms of building soc and DR price fluctuations. Specifically,  we generally see an increasing soc at low price and a decreasing soc at high price. Moreover, we see that  the soc has been maintained within the range of $[-1, 1]$. This actually indicates that the building has leveraged the operational flexibility of multi-zone HVAC system in response of dynamic electricity price. 

\section{Conclusion}
This paper developed a unified virtual battery (VB) modeling framework  for characterizing the operating flexibility of both single-zone  and multi-zone HVAC systems. The proposed VB models were  established from the widely-used RC formulations with theoretic guarantee. The proposed VB models show high interpretability and enable flexible buildings to function like virtual batteries. 
 We  demonstrated the effectiveness of  VB models for characterizing building thermal dynamics  and their near-optimal performance  for building  demand response (DR) applications. 
The  VB models are applicable to diverse building-level energy management or scheduling tasks and provide a solid model foundation for unlocking the substantial operating flexibility of buildings for supporting modern power grid operation.  Several promising future directions have been identified. First, it is interesting to  integrate the   VB model into the deep reinforcement learning framework and explore whether a low-order model can accelerate the learning rate and lead to better building energy management strategies. Second, the  VB model can be further extended for aggregating the operating flexibility of building clusters for providing  grid service at the microgrid or city level.

{\appendices

 \section*{Appendix A}
 For single-zone HVAC, the characterization state is 
 \begin{equation} \label{eq:AC_soz}
 	{\rm soz}_{n}(k) = \frac{T_n^{\rm max} - T_{n}(k)}{2\delta_n}, \quad \forall k \in \mathcal{K}.
 \end{equation}

 When the HVAC unit is operated under \texttt{Baseline control}, i.e.,  maintaining the zone temperature at the upper temperature limit, we have the following steady-state equations according to  \eqref{eq:AC_dynamics}: 
 \begin{equation} \label{eq:AC_dynamics_base}
 	T_{n}^{\rm max}(k\!+\!1) \!\!=\! a_n T_{n}^{\rm max}(k)\!-\!b_n q_{n}^{\rm HVAC,  base}(k) + d_{n}(k), \forall k \in \mathcal{K}.
 \end{equation}
 
 By subtracting \eqref{eq:AC_dynamics} from \eqref{eq:AC_dynamics_base}, and dividing $2\delta_n$, we have  
 \begin{equation}  \label{eq:equation}
 	\begin{split}
 		& \frac{T_{n}^{\rm max}(k+1) - T_{n}(k+1)}{2 \delta_n}\!\!=\! a_n \frac{T_{n}^{\rm max}(k) - T_{n}(k+1)}{2 \delta_n}\! \\
 		& \quad \quad \quad \quad + \!b_n \frac{q_{n}^{\rm HVAC, base}(k) -q_{n}^{\rm HVAC}(k)}{2\delta_n}, \forall k \in \mathcal{K}.
 	\end{split}
 \end{equation}
 By substituting \eqref{eq:AC_dynamics_base} into \eqref{eq:equation}, and combine with the HVAC operating limits and thermal comfort constraints, we obtain the following VB model for single-zone HVAC system:
 \begin{equation} 
 	\begin{split}
 		& \textbf{VB~model~for-single-zone~HVAC:} \\
 		& \begin{cases}
 			{\rm soz}_{n}(k\!+\!1)\!\!=\!a_n {\rm soz}_{n}(k)\!\!+\!\!P^{\rm ch/dis}_{n}(k), \\
 			q_{n}^{\rm HVAC, \min}(k) \leq q_{n}^{\rm HVAC}(k) \leq q_{n}^{\rm HVAC, \max}(k),\\
 			0 \leq {\rm soz}_n(k) \leq 1, \forall k \in \mathcal{K}. 
 		\end{cases}
 	\end{split}
 \end{equation}
 where $P^{\rm ch/dis}_{n}(k) = b_n/(2\delta_n) \left(q_{n}^{\rm HVAC}(k) \!-\! q_{n}^{\rm HVAC, base}(k) \right)$ denotes the net charging energy of the VB model at time step $k$. The  energy consumption of the HVAC  at time step $k$ is $q_{n}^{\rm HVAC}(k) \Delta k$, which can be equivalently expressed as $ Q^{\rm tol}_n(k) = \big(2 P^{\rm ch/dis}_{n}(k)\delta_n/b_n + q_{n}^{\rm HVAC, base}(k) \big) \Delta k $ with the VB model.
 
\section*{Appendix B}
 For multi-zone HVAC, the characterization states are 
\begin{equation} \label{eq:AC_soz}
	{\rm soz}_{n, i}(k) = \frac{T_{n, i}^{\rm max} - T_{n, i}(k)}{2\delta_{n, i}}, \quad \forall i \in \mathcal{I}_n,  k \in \mathcal{K}.
\end{equation}

When the system is operating under \texttt{Baseline control}: maintaining the zone  to the upper comfortable temperature limits $T_{n,i}^{\rm max} \triangleq T_{n,i}^{\rm set} + \delta_{n, i}$, we have the  following steady-state equations according to \eqref{eq:optimize}: 
\begin{equation} \label{eq:new_baseline}
	\begin{split}
		& \T^{\rm max}_n(k\!+\!1) \!\!=\!\! \A_n \T^{\rm max}_n(k) \!+\!\B_n \q^{\rm base}_n(k) \!+\! \aaa^{\rm out}_n \T^{\rm out}(k) \!\!+\!\! \dd_n(k), \\
		& \q^{\rm base}_n(k) \!=\! c_p \m_n(k)\left(\T^{\rm max}_n(k) - \mathbf{1}_{N_n} T^{\rm sup}_n \right), \forall k \in \mathcal{K}.
	\end{split}
\end{equation}

By subtracting \eqref{eq:optimize} from \eqref{eq:new_baseline},  and dividing $2\delta_{n, i}$ for each of the stacked equations, we have 
\begin{equation} 
	\label{eq:zone_2}
	\begin{split}
		& \frac{T_{n, i}^{\rm max}(k+1) \!\!-\! T_{n, i}(k)}{2\delta_{n, i}}\!=\! a_{n, ii} \frac{T_{n, i}^{\rm max}\!-\! T_{n, i}(k)}{2\delta_{n, i}} \!\! \\
		& +\!\!\!\! \sum_{j \in \mathcal{I}_n(i)}\!\! \frac{\delta_{n, j}}{ \delta_{n, i}}\frac{T_{n, j}^{\rm max} \!-\! T_{n, j}(k)}{2\delta_{n, j}}\!+\!\!\! \frac{b_{n, i}}{2\delta_{n, i}}(q_{n, i}(k) \!-\! q_{n,i}^{\rm base}(k)), \\
		& \quad \quad  \quad \quad \quad  \quad \forall i \in \mathcal{I}_n, k \in \mathcal{K}. 
	\end{split}
\end{equation}

By substituting  \eqref{eq:new_soz} into \eqref{eq:zone_2}, we further have 
\begin{equation} \label{eq:zone_soz1}
	\begin{split}
		{\rm soz}_{n, i}&(k + 1)\!\! =\!\! a_{n, ii} {\rm soz}_{n, i}(k) \!\! +\!\!\!\!\sum_{j \in \mathcal{I}_n(i)}\!\!\! a_{n, ij} \frac{\delta_{n, j}}{\delta_{n, i}} {\rm soz}_{n, j}(k) \\
		&+ \frac{b_{n, i}}{2\delta_{n, i}} (q_{n, i}(k) - q_{n, i}^{\rm base}(k)), ~\forall i \in \mathcal{I}_n, k \in \mathcal{K}. 
	\end{split}
\end{equation}

By stacking \eqref{eq:zone_soz1}  across all zones, we obtain 
the  VB models for multi-zone HVAC systems: 
\begin{equation} \label{eq:soz_equations_1}
	\begin{split}
		& {\rm \textbf{Zone~VB~models for multi-zone HVAC system:}}\\
		& \begin{cases}
			\soz_n(k\!+\!1) \!\!=\!\! \tilde{\A}_n \soz_n(k) \!+\! \tilde{\B}_n\!\!\left(\q_n(k) \!-\! \q_n^{\rm base}(k) \right) \\
			\mathbf{0}_n \leq \soz_n(k) \leq \mathbf{1}_n, \\
			\q_n^{\min}(k) \leq \q_n(k) \leq \q_n^{\max}(k), \quad \forall k \in \mathcal{K}. 
		\end{cases}
	\end{split}
\end{equation}
where we have 
\begin{equation}
	\setlength{\arraycolsep}{2pt} 
	\renewcommand{\arraystretch}{0.9} 
	\begin{aligned}
		& \tilde{\A}_n = 
		\begin{bmatrix}
			a_{n,11} & a_{n,12} \frac{\delta_{n,2}}{\delta_{n,1}} & \dots & a_{n,1N_n} \frac{\delta_{n,N_n}}{\delta_{n,1}} \\
			a_{n,21}\frac{\delta_{n,1}}{\delta_{n,2}} & a_{n,22} & \dots & a_{n,2N_n} \frac{\delta_{n,N_n}}{\delta_{n,2}}\\
			\vdots & \vdots & \ddots & \vdots \\
			a_{n,N_n1}\frac{\delta_{n,1}}{\delta_{n,N_n}} & a_{n,N_n2} \frac{\delta_{n,2}}{\delta_{n,N_n}} & \dots & a_{n,N_nN_n}
		\end{bmatrix} \in R^{I_n \times I_n}, \\[2mm]  
		& \tilde{\B}_n = 
		\mathrm{diag}\left(\frac{b_{n,1}}{2\delta_{n,1}}, \dots, \frac{b_{n,N_n}}{2\delta_{n,N_n}}\right) \in R^{I_n \times I_n}, \\
		& \mathbf{0}_n = [0, 0, \cdots, 0] \in R^{I_n}, \mathbf{1}_n = [1, 1, \cdots, 1] \in R^{I_n}. 
	\end{aligned}
	\renewcommand{\arraystretch}{1} 
	\setlength{\arraycolsep}{5pt} 
\end{equation}

Subsequently, by following the proposed multi-zone VB aggregation framework in Section IV-D, we  can obtain the aggregated VB model  for building multi-zone HVAC systems: 
\begin{equation} \label{eq:new_overall_VB_model}
	\begin{split}
		& {\rm \textbf{Aggregated VB model for multi-zone HVAC:}} \\
		& \begin{cases}
			%
			\soc_n(k+1) = \alpha_n \soc_n(k) \!+\! P_n^{\rm ch/dis}(k), \\
			P^{\rm ch/dis}_n(k) \!\geq\! \beta_n^{\min}(k) Q_n(k) \!-\! \w_n^{\mathsf T} \tilde{\B}_n\q^{\rm base}_n(k), \\
			P^{\rm ch/dis}_n(k)\!\leq\! \beta_n^{\max}(k) Q_n(k) \!-\! \w_n^{\mathsf T} \tilde{\B}_n\q^{\rm base}_n(k), \\
			Q_n^{\min}(k) \leq Q_n(k) \leq Q_n^{\max}(k), \\
			0 \leq \soc_n(k) \leq 1, \quad \quad  \forall k \in \mathcal{K}. \\
		\end{cases}
	\end{split}
\end{equation}
The energy consumption model can be established using the same data-driven approach introduced in \eqref{eq:energy_model} with the VB model.

\bibliographystyle{ieeetr}  
\bibliography{reference.bib}

\begin{thebibliography}{10}

\bibitem{IEA_Energy_Efficiency_2025}
{International Energy Agency}, ``Energy efficiency 2025,'' 2025.
\newblock Licence: CC BY 4.0.

\bibitem{wijesuriya2025enhancing}
S.~Wijesuriya, R.~A. Kishore, M.~Mitchell, and C.~Booten, ``Enhancing
  energyplus capabilities to model dynamic building envelopes using python
  plugin,'' {\em Energy and Buildings}, p.~115776, 2025.

\bibitem{GEB_WholeBuilding_2019}
A.~Roth {\em et~al.}, ``Grid‑interactive efficient buildings:
  Whole‑building controls, sensors, modeling, and analytics,'' Tech. Rep.
  NREL/TP--5500-75478; DOE/GO--102019-5230, U.S. Department of Energy, Building
  Technologies Office; National Renewable Energy Laboratory, 2019.

\bibitem{kou2021model}
X.~Kou, Y.~Du, F.~Li, H.~Pulgar-Painemal, H.~Zandi, J.~Dong, and M.~M. Olama,
  ``Model-based and data-driven hvac control strategies for residential demand
  response,'' {\em IEEE Open Access Journal of Power and Energy}, vol.~8,
  pp.~186--197, 2021.

\bibitem{anuntasethakul2021design}
C.~Anuntasethakul and D.~Banjerdpongchai, ``Design of supervisory model
  predictive control for building hvac system with consideration of peak-load
  shaving and thermal comfort,'' {\em IEEE access}, vol.~9, pp.~41066--41081,
  2021.

\bibitem{qureshi2018hierarchical}
F.~A. Qureshi and C.~N. Jones, ``Hierarchical control of building hvac system
  for ancillary services provision,'' {\em Energy and Buildings}, vol.~169,
  pp.~216--227, 2018.

\bibitem{drgovna2020all}
J.~Drgo{\v{n}}a, J.~Arroyo, I.~C. Figueroa, D.~Blum, K.~Arendt, D.~Kim, E.~P.
  Oll{\'e}, J.~Oravec, M.~Wetter, D.~L. Vrabie, {\em et~al.}, ``All you need to
  know about model predictive control for buildings,'' {\em Annual reviews in
  control}, vol.~50, pp.~190--232, 2020.

\bibitem{energyplus}
{U.S. Department of Energy and Lawrence Berkeley National Laboratory},
  ``Energyplus energy simulation software.'' \url{https://energyplus.net/},
  2025.
\newblock Version X.Y.Z (replace with the version you used).

\bibitem{trnsys}
S.~A. Klein, W.~A. Beckman, J.~A. Duffie, {\em et~al.}, ``Trnsys: A transient
  system simulation program.'' \url{http://sel.me.wisc.edu/trnsys}, 2017.
\newblock Solar Energy Laboratory, University of Wisconsin–Madison.

\bibitem{smarra2018data}
F.~Smarra, A.~Jain, T.~De~Rubeis, D.~Ambrosini, A.~D’Innocenzo, and
  R.~Mangharam, ``Data-driven model predictive control using random forests for
  building energy optimization and climate control,'' {\em Applied energy},
  vol.~226, pp.~1252--1272, 2018.

\bibitem{lu2021data}
S.~Lu, W.~Gu, S.~Ding, S.~Yao, H.~Lu, and X.~Yuan, ``Data-driven aggregate
  thermal dynamic model for buildings: A regression approach,'' {\em IEEE
  Transactions on Smart Grid}, vol.~13, no.~1, pp.~227--242, 2021.

\bibitem{deng2010building}
K.~Deng, P.~Barooah, P.~G. Mehta, and S.~P. Meyn, ``Building thermal model
  reduction via aggregation of states,'' in {\em Proceedings of the 2010
  American Control Conference}, pp.~5118--5123, IEEE, 2010.

\bibitem{maasoumy2011model}
M.~Maasoumy, A.~Pinto, and A.~Sangiovanni-Vincentelli, ``Model-based
  hierarchical optimal control design for hvac systems,'' in {\em Dynamic
  systems and control conference}, vol.~54754, pp.~271--278, 2011.

\bibitem{yang2021distributed}
Y.~Yang, S.~Srinivasan, G.~Hu, and C.~J. Spanos, ``Distributed control of
  multizone hvac systems considering indoor air quality,'' {\em IEEE
  Transactions on Control Systems Technology}, vol.~29, no.~6, pp.~2586--2597,
  2021.

\bibitem{yang2020hvac}
Y.~Yang, G.~Hu, and C.~J. Spanos, ``Hvac energy cost optimization for a
  multizone building via a decentralized approach,'' {\em IEEE Transactions on
  Automation Science and Engineering}, vol.~17, no.~4, pp.~1950--1960, 2020.

\bibitem{cui2024data}
X.~Cui, S.~Liu, G.~Ruan, and Y.~Wang, ``Data-driven aggregation of thermal
  dynamics within building virtual power plants,'' {\em Applied Energy},
  vol.~353, p.~122126, 2024.

\bibitem{cui2025dimension}
X.~Cui, Y.~Wang, and B.~Xu, ``Dimension-reduced optimization of multi-zone
  thermostatically controlled loads,'' {\em IEEE Transactions on Smart Grid},
  2025.

\bibitem{hao2017optimal}
H.~Hao, D.~Wu, J.~Lian, and T.~Yang, ``Optimal coordination of building loads
  and energy storage for power grid and end user services,'' {\em IEEE
  Transactions on Smart Grid}, vol.~9, no.~5, pp.~4335--4345, 2017.

\bibitem{zhao2017geometric}
L.~Zhao, W.~Zhang, H.~Hao, and K.~Kalsi, ``A geometric approach to aggregate
  flexibility modeling of thermostatically controlled loads,'' {\em IEEE
  Transactions on Power Systems}, vol.~32, no.~6, pp.~4721--4731, 2017.

\bibitem{mukhi2025aggregate}
K.~Mukhi and A.~Abate, ``Aggregate flexibility of thermostatically controlled
  loads using generalized polymatroids,'' {\em arXiv preprint
  arXiv:2504.00484}, 2025.

\bibitem{hou2024privacy}
Z.~Hou, S.~Lu, Y.~Xu, H.~Qiu, W.~Gu, Z.~Y. Dong, and S.~Ding,
  ``Privacy-preserved aggregate thermal dynamic model of buildings,'' {\em IEEE
  Transactions on Smart Grid}, vol.~15, no.~6, pp.~5653--5664, 2024.

\bibitem{song2017thermal}
M.~Song, C.~Gao, H.~Yan, and J.~Yang, ``Thermal battery modeling of inverter
  air conditioning for demand response,'' {\em IEEE Transactions on Smart
  Grid}, vol.~9, no.~6, pp.~5522--5534, 2017.

\bibitem{song2019hierarchical}
M.~Song, W.~Sun, Y.~Wang, M.~Shahidehpour, Z.~Li, and C.~Gao, ``Hierarchical
  scheduling of aggregated tcl flexibility for transactive energy in power
  systems,'' {\em IEEE Transactions on Smart Grid}, vol.~11, no.~3,
  pp.~2452--2463, 2019.

\bibitem{zhu2025multi}
X.~Zhu, P.~Wang, N.~Li, and W.~Yan, ``Multi-period optimal scheduling of
  building loads based on accurate virtual battery model,'' {\em Energy and
  Buildings}, vol.~327, p.~115046, 2025.

\bibitem{abbas2022evaluation}
A.~Abbas, R.~Ariwoola, B.~Chowdhury, S.~Kamalasadan, and Y.~Lin, ``Evaluation
  of equivalent battery model representations for thermostatically controlled
  loads in commercial buildings,'' in {\em 2022 IEEE Industry Applications
  Society Annual Meeting (IAS)}, pp.~1--6, IEEE, 2022.

\bibitem{guo2019identification}
Z.~Guo, A.~R. Coffman, J.~Munk, P.~Im, and P.~Barooah, ``Identification of
  aggregate building thermal dynamic model and unmeasured internal heat load
  from data,'' in {\em 2019 IEEE 58th Conference on Decision and Control
  (CDC)}, pp.~2958--2963, IEEE, 2019.

\bibitem{guo2021aggregation}
Z.~Guo, A.~R. Coffman, J.~Munk, P.~Im, T.~Kuruganti, and P.~Barooah,
  ``Aggregation and data driven identification of building thermal dynamic
  model and unmeasured disturbance,'' {\em Energy and Buildings}, vol.~231,
  p.~110500, 2021.

\bibitem{qiu2023federated}
D.~Qiu, J.~Xue, T.~Zhang, J.~Wang, and M.~Sun, ``Federated reinforcement
  learning for smart building joint peer-to-peer energy and carbon allowance
  trading,'' {\em Applied Energy}, vol.~333, p.~120526, 2023.

\bibitem{lin2015experimental}
Y.~Lin, P.~Barooah, S.~Meyn, and T.~Middelkoop, ``Experimental evaluation of
  frequency regulation from commercial building hvac systems,'' {\em IEEE
  Transactions on Smart Grid}, vol.~6, no.~2, pp.~776--783, 2015.

\bibitem{wang2017distributed}
Z.~Wang, G.~Hu, and C.~J. Spanos, ``Distributed model predictive control of
  bilinear hvac systems using a convexification method,'' in {\em 2017 11th
  Asian control conference (ASCC)}, pp.~1608--1613, IEEE, 2017.

\bibitem{nweye2025citylearn}
K.~Nweye, K.~Kaspar, G.~Buscemi, T.~Fonseca, G.~Pinto, D.~Ghose, S.~Duddukuru,
  P.~Pratapa, H.~Li, J.~Mohammadi, {\em et~al.}, ``Citylearn v2:
  energy-flexible, resilient, occupant-centric, and carbon-aware management of
  grid-interactive communities,'' {\em Journal of Building Performance
  Simulation}, vol.~18, no.~1, pp.~17--38, 2025.

\end{thebibliography}

\end{document}